\documentclass[aps,prd,superscriptaddress,showkeys,showpacs,nofootinbib,10pt,twocolumn,longbibliography,floatfix]{revtex4-2}

\usepackage[english]{babel}

\usepackage{color}
\usepackage{graphicx}
\usepackage{amsmath,amssymb,empheq}
\graphicspath{{./figures/}}
\usepackage{appendix}
\usepackage{comment}

\allowdisplaybreaks

\usepackage[breaklinks=true]{hyperref}
\hypersetup{
	colorlinks=true,
	linkcolor=blue,
	filecolor=magenta,      
	urlcolor=blue,
	citecolor=red,
}

\usepackage{orcidlink}

\usepackage{mathrsfs}
\usepackage[normalem]{ulem}

\usepackage{placeins}
\usepackage[utf8]{inputenc}

\input epsf
\usepackage{mathtools}
\usepackage{amsfonts}
\usepackage{upgreek}
\usepackage{latexsym}
\usepackage{afterpage}
\usepackage{bm}

\usepackage{siunitx}
\usepackage{multirow}

\usepackage{stackengine}
\usepackage{scalerel}
\usepackage{xcolor}

\begin{document}

{\hfill MS-TP-26-28}
\title{Gauge-field energy density in inflationary kinetic-coupling models:\\
QFT regularization versus horizon cutoffs}

\author{A.O.~Naidenko\,\orcidlink{0009-0004-2129-8982}}
\email{naydenko\_a@knu.ua}
\affiliation{Physics Faculty, \href{https://ror.org/02aaqv166}{Taras Shevchenko National University of Kyiv}, 64/13, Volodymyrska Street, 01601 Kyiv, Ukraine}

\author{O.O.~Sobol\,\orcidlink{0000-0002-6300-3079}}
\email{oleksandr.sobol@knu.ua}
\affiliation{Institute for Theoretical Physics, \href{https://ror.org/00pd74e08}{University of M\"{u}nster}, Wilhelm-Klemm-Stra{\ss}e 9, 48149 M\"{u}nster, Germany}
\affiliation{Physics Faculty, \href{https://ror.org/02aaqv166}{Taras Shevchenko National University of Kyiv}, 64/13, Volodymyrska Street, 01601 Kyiv, Ukraine}
	
\author{S.I.~Vilchinskii\,\orcidlink{0000-0002-9294-9939}}
\email{stanislav.vilchinskii@cern.ch}
\affiliation{D\'{e}partement de Physique Th\'{e}orique  and Center for Astroparticle Physics, \href{https://ror.org/01swzsf04}{Universit\'{e} de Gen\`{e}ve},  24 quai  Ernest Ansermet, 1211 Gen\`{e}ve 4, Switzerland}
% \affiliation{Niels Bohr Institute, \href{https://ror.org/035b05819}{University of Copenhagen}, Blegdamsvej 17, 2010 Copenhagen, Denmark}
\affiliation{Physics Faculty, \href{https://ror.org/02aaqv166}{Taras Shevchenko National University of Kyiv}, 64/13, Volodymyrska Street, 01601 Kyiv, Ukraine}

\date{\today}

%%%%%%%%%%%%%%%%%%%%%%%%%%%%%%%%%%%%%%%%%%%%%%%%%%%%%%%%%%%%%%%%%%%%%%%%%%
\begin{abstract}
    Coupling gauge fields to the inflaton can amplify their quantum fluctuations during inflation, with potential applications to a range of cosmological phenomena. A central challenge is to define finite and physically meaningful observables, such as the gauge-field energy density. In this work, we study an Abelian gauge field kinetically coupled to the inflaton through $f^2(\phi)F_{\mu\nu}F^{\mu\nu}$ in a de~Sitter background with $f\propto a^{\gamma_1}$, and compare conventional quantum-field-theoretic regularization with horizon-cutoff prescriptions. Using a hard physical-momentum cutoff and dimensional regularization, we derive the ultraviolet divergences in closed form and fix the remaining one-parameter finite-part ambiguity by requiring the coupling-induced total energy density to be nonnegative. We then examine five horizon-cutoff prescriptions based on characteristic scales of the mode equations. Their predictions depend sensitively on the choice of horizon scale and, for weak gauge-field production, can deviate substantially from the regularized result; for the adopted finite-part prescription, the minimal horizon agrees best with the regularized total energy density. As $|\gamma_1|\to 2$, where infrared modes dominate, the relative deviation of all prescriptions from the regularized result vanishes, independently of the finite-part prescription. Outside this regime, the separate electric and magnetic contributions, which can become negative after regularization, are not reliably reproduced by any of the horizon-cutoff prescriptions.
\end{abstract}
%%%%%%%%%%%%%%%%%%%%%%%%%%%%%%%%%%%%%%%%%%%%%%%%%%%%%%%%%%%%%%%%%%%%%%%%%%

\maketitle

%%%%%%%%%%%%%%%%%%%%%%%%%%%%%%%%%%%%%%%%%%%%%%%%%%%%%%%%%%%%%%%%%%%%%%%%%%
\section{Introduction}
\label{sec:intro} 

Models of cosmic inflation~\cite{Starobinsky:1980te,Guth:1980zm,Linde:1981mu,Albrecht:1982wi,Linde:1983gd} that involve an additional Abelian gauge field coupled to the inflaton are interesting from both theoretical and observational points of view and exhibit a rich phenomenology; see Refs.~\cite{Kandus:2010nw,Durrer:2013pga,Maleknejad:2012fw,Subramanian:2015lua,Vachaspati:2020blt} for reviews. Such models have been investigated, among other applications, as mechanisms for generating primordial magnetic fields~\cite{Turner:1987bw,Ratra:1991bn,Garretson:1992vt,Martin:2007ue,Demozzi:2009fu,Durrer:2010mq,Ferreira:2013sqa,Caprini:2014mja,Fujita:2015iga,Adshead:2016iae,Sharma:2017eps,Sobol:2020lec,Gorbar:2021zlr}, for baryogenesis~\cite{Anber:2015yca,Jimenez:2017cdr,Kamada:2020bmb}, and for gauge preheating at the end of inflation~\cite{Deskins:2013dwa,Adshead:2015pva,Adshead:2018doq,Cuissa:2018oiw}, as well as for producing distinctive signatures in the cosmic microwave background (CMB)~\cite{Barnaby:2010vf,Barnaby:2011vw,Meerburg:2012id,Watanabe:2010fh,Soda:2012zm,Bartolo:2012sd,Planck:2019kim} and the stochastic gravitational-wave background~\cite{Sorbo:2011rz,Barnaby:2011qe,Namba:2015gja,Domcke:2016bkh,Garcia-Bellido:2016dkw,Adshead:2019igv,Bastero-Gil:2022fme,Corba:2024tfz,Sharma:2024nfu,vonEckardstein:2025oic,vonEckardstein:2025elq,Teuscher:2025jhq}. Primordial magnetic fields generated in this way could seed the magnetic fields observed in galaxies, galaxy clusters, and cosmic filaments~\cite{Kronberg:1993vk,Bernet:2008qp,Vallee:2011zz,Feretti:2012vk,Vacca:2018rta,Carretti:2022fqk} and account for the intergalactic magnetic fields suggested by gamma-ray observations of blazars~\cite{Neronov:2010gir,Tavecchio:2010mk,Ando:2010rb,Dolag:2010ni,Taylor:2011bn,Vovk:2011aa,MAGIC:2022piy} (see Ref.~\cite{AlvesBatista:2021sln} for a review). Despite extensive investigation over more than three decades, however, fundamental questions remain concerning the precise treatment of the quantum gauge field and, in particular, the definition of its physical energy density during inflation.
 
Gauge-field production during inflation generally requires a breaking of the conformal invariance of Maxwell theory~\cite{Parker:1968mv,Turner:1987bw,Dolgov:1993vg}. This can be achieved, for example, through a time-dependent kinetic coupling of the form $f^2(\phi)F_{\mu\nu}F^{\mu\nu}$~\cite{Ratra:1991bn,Giovannini:2001nh,Bamba:2003av,Martin:2007ue,Demozzi:2009fu,Kanno:2009ei,Vilchinskii:2017qul,Sobol:2018djj,Talebian:2020drj,Sobol:2020lec,Maity:2021qps,Sasaki:2022rat,Dimopoulos:2024jnv,Li:2025crs}, through a pseudoscalar coupling $\phi F_{\mu\nu}\tilde F^{\mu\nu}$~\cite{Turner:1987bw,Garretson:1992vt,Anber:2006xt,Durrer:2010mq,Caprini:2014mja,Ng:2014lyb,Fujita:2015iga,Adshead:2016iae,Sobol:2019xls,Shtanov:2019civ,Gorbar:2021zlr,Cecchini:2023bqu}, or through nonminimal couplings to gravity~\cite{Turner:1987bw,Bamba:2008ja,Savchenko:2018pdr,Bamba:2020qdj,Sobol:2021jag,Durrer:2022emo}. The time dependence of these couplings modifies the evolution of the gauge-field modes and can lead to substantial amplification of quantum fluctuations. In axion-like models, for example, one helicity can undergo a tachyonic instability, resulting in highly efficient production of gauge quanta~\cite{Garretson:1992vt,Anber:2006xt,Anber:2009ua,Barnaby:2010vf}. The amplified gauge-field fluctuations can subsequently source scalar and tensor perturbations~\cite{Seery:2008ms,Caldwell:2011ra,Motta:2012rn,Barnaby:2012xt,Jain:2012ga,Ferreira:2014zia,Ferreira:2015omg,Papageorgiou:2019ecb,Durrer:2024ibi} and thereby leave potentially observable imprints, ranging from the CMB and gravitational-wave signatures mentioned above to primordial black holes~\cite{Linde:2012bt,Garcia-Bellido:2016dkw,Ozsoy:2023ryl}; in magnetogenesis scenarios, the induced curvature perturbations also yield stringent constraints~\cite{Suyama:2012wh,Ringeval:2013hfa,Fujita:2013qxa,Nurmi:2013gpa,Fujita:2014sna,Ferreira:2014hma}.
 
The efficiency of gauge-field production is controlled by the evolution and strength of the inflaton--gauge-field coupling and by the inflationary background. When the produced gauge fields become sufficiently energetic, their contribution to the inflaton and Friedmann equations can no longer be neglected~\cite{Demozzi:2009fu,Kanno:2009ei,Sharma:2017eps,Talebian:2020drj}. Their backreaction can then modify the evolution of the inflaton and, consequently, the subsequent production of gauge fields, as studied for both kinetic~\cite{Watanabe:2009ct,Kanno:2009ei,Sobol:2018djj,Sobol:2020lec,Durrer:2023rhc,Lysenko:2025pse} and axion-like couplings~\cite{Anber:2009ua,Cheng:2015oqa,Notari:2016npn,Domcke:2020zez,Gorbar:2021rlt,Peloso:2022ovc,Garcia-Bellido:2023ser,vonEckardstein:2023gwk,Galanti:2024jhw}. Self-consistent descriptions of this regime have been developed using, for example, the gradient-expansion formalism~\cite{Sobol:2019xls,Sobol:2020lec,Gorbar:2021rlt,Durrer:2023rhc,vonEckardstein:2023gwk,Lysenko:2025pse,vonEckardstein:2025jug,Lysenko:2025sdo}, the stochastic formalism~\cite{Fujita:2022fit}, perturbative treatments of inhomogeneities~\cite{Domcke:2023tnn}, and lattice simulations~\cite{Caravano:2021bfn,Caravano:2022epk,Figueroa:2023oxc,Caravano:2024xsb,Sharma:2024nfu,Figueroa:2024rkr}. In all cases, a self-consistent treatment of gauge-field production requires a reliable determination of the gauge-field energy density and of its contribution to the background dynamics.
 
This leads to a fundamental issue: the gauge-field energy density is a quantum expectation value and, in general, is ultraviolet (UV) divergent. More precisely, the energy density is obtained from the expectation value of the energy--momentum tensor, whose mode decomposition contains contributions from arbitrarily large momenta. Consequently, a prescription is required to regulate the UV divergence and to identify the finite contribution associated with the physical state of the field. This problem is familiar from quantum field theory (QFT) in curved spacetime~\cite{Birrell:1982ix,Parker:2009uva}, where expectation values of composite operators such as the energy--momentum tensor require regularization and, ultimately, renormalization, for instance by adiabatic subtraction~\cite{Parker:1974qw,Chu:2016kwv,Maranon-Gonzalez:2023efu} or point splitting~\cite{Bunch:1978yq}; see Ref.~\cite{Negro:2024bbf} for a recent review in the context of primordial observables. The backreaction of the renormalized energy--momentum tensor on the cosmological background was already considered in early work~\cite{Fischetti:1979ue,Starobinsky:1980te}.
 
For inflationary gauge-field production, an additional conceptual issue arises from the strong scale dependence of the amplification. Modes well inside the horizon can remain close to their vacuum behavior, whereas modes around and beyond characteristic horizon scales can be strongly affected by the time-dependent kinetic coupling. This motivates prescriptions in which the momentum integral defining the energy density is separated at a horizon-related scale, with the modes affected by gauge-field production treated as the physically relevant contribution~\cite{Demozzi:2009fu,Sobol:2018djj,Sobol:2019xls,Sobol:2020lec,Gorbar:2021rlt,Fujita:2022fit,Durrer:2023rhc,vonEckardstein:2023gwk,Durrer:2024ibi,Lysenko:2025pse,Lysenko:2025sdo,Teuscher:2025jhq}. Such horizon-based prescriptions provide a physically motivated way of isolating the amplified part of the spectrum, but they represent a different separation of modes from that introduced by a UV cutoff and, as noted in Refs.~\cite{Durrer:2024ibi,Lysenko:2025sdo}, do not amount to a rigorous renormalization.
 
A related analysis was carried out by Ballardini et al.~\cite{Ballardini:2019rqh}, who studied the energy--momentum tensor of gauge fields coupled to a pseudoscalar inflaton through an axionlike coupling. Assuming an approximately constant inflaton velocity, they derived analytical expressions for the divergent and finite contributions to the gauge-field energy density and identified the associated counterterms using an adiabatic subtraction. Refinements of adiabatic renormalization that modify the subtraction in the infrared have since been proposed, based either on a physically motivated infrared cutoff~\cite{Animali:2022lig} or on an additional physical scale~\cite{Ferreiro:2023uvr}, while the impact of such subtractions on other inflationary observables, most notably the primordial power spectra, has been debated~\cite{Agullo:2009vq,Durrer:2009ii,Corba:2022ugu,Pla:2024xsv}. These works highlight the importance of a systematic treatment of UV divergences during inflation. The present study addresses a complementary question: how do conventional QFT regularization and horizon-cutoff prescriptions compare for a gauge field kinetically coupled to the inflaton?
 
We investigate two approaches to computing the gauge-field energy density. The first employs conventional QFT regularization techniques: hard-cutoff regularization, in which the momentum integrals are regulated by imposing a cutoff on the physical momentum of the gauge field, and dimensional regularization, in which the momentum integral is analytically continued to a noninteger number of spatial dimensions. The second employs explicit horizon scales to separate modes substantially affected by the kinetic coupling from those that remain effectively unaffected. We examine five different horizon prescriptions and study how the resulting energy densities depend on the choice of horizon scale.
 
Our comparison reveals significant differences between the conventional QFT and horizon-based prescriptions, with the latter generally predicting enhanced gauge-field energy densities and exhibiting qualitatively different behavior over much of the parameter range considered. In particular, the horizon-cutoff results depend sensitively on the choice of horizon scale. Nevertheless, all five horizon prescriptions considered here approach the regularized QFT result in the regime of strong gauge-field production, as $|\gamma_1|$ approaches $2$, while the uncertainty associated with variations of the horizon scale becomes smaller. These findings demonstrate that the prescription used to define the gauge-field energy density can significantly affect its predicted value in the regime of weak gauge-field production, whereas the different prescriptions converge in the infrared-dominated regime of strong gauge-field production, where backreaction on the inflationary background may eventually become important.

\textbf{The paper is organized as follows.} In Sec.~\ref{sec:model}, we introduce the inflationary model with a kinetically coupled Abelian gauge field and describe basic equations governing the gauge-field dynamics and its energy density. In Sec.~\ref{sec:ultraviolet_regularization}, we develop the hard-cutoff and dimensional-regularization approaches to the gauge-field energy density, derive the corresponding UV structures, compare their results, and discuss the prescription used to fix the finite part. In Sec.~\ref{sec:approach_comparison}, we consider five horizon-cutoff prescriptions and compare their results with the regularized QFT result of Sec.~\ref{sec:ultraviolet_regularization}, highlighting the differences between the corresponding definitions of the gauge-field energy density. Finally, Sec.~\ref{sec:conclusions} summarizes our main results and conclusions. Technical details of the Hankel-function integrals and their asymptotic expansion are collected in Appendix~\ref{app:Hankel-integrals}. A component-wise comparison of the electric and magnetic energy densities is presented in Appendix~\ref{app:comparison-components}.

\textbf{Notations.} Throughout this work we adopt natural units, setting $\hbar = c = 1$, and denote the reduced Planck mass by $M_{\mathrm{P}} = \SI{2.43e18}{\giga\electronvolt}$. The inflationary background is taken to be homogeneous and isotropic, described by the spatially flat Friedmann--Lema\^{i}tre--Robertson--Walker (FLRW) metric. In terms of cosmic time $t$ or conformal time $\eta$, the line element can be written as
\begin{equation}
\label{eq:FLRW-metric}
    ds^2\! =\! g_{\mu\nu}\,\mathrm{d}x^\mu \mathrm{d}x^\nu\! = \mathrm{d}t^2 \!- \!a^2(t)\,\mathrm{d}\bm{x}^2 %\nonumber\\
    = a^2(\eta)\left[\mathrm{d}\eta^2 \!-\! \mathrm{d}\bm{x}^2\right],
\end{equation}
where $a(t)$ denotes the scale factor. Derivatives with respect to cosmic time $t$ are indicated by overdots, while those with respect to conformal time $\eta$ are denoted by primes. The Hubble parameter is defined as $H \equiv \dot{a}/a$.

%%%%%%%%%%%%%%%%%%%%%%%%%%%%%%%%%%%%%%%%%%%%%%%%%%%%%%%%%%%%%%%%%%%%%%%%%%
\section{Model and gauge-field dynamics}
\label{sec:model}

\subsection{Kinetic coupling model}
\label{subsec:background}

We consider a $U(1)$ gauge field $A_\mu$ kinetically coupled to a scalar inflaton field $\phi$. The action describing the system is
\begin{equation}
S = \int \mathrm{d}^4 x \sqrt{-g}
\bigg[
\frac{1}{2}\partial_\mu\phi \partial^\mu\phi
- V(\phi)
- \frac{f^2(\phi)}{4}F_{\mu\nu}F^{\mu\nu}
\bigg],
\end{equation}
where $F_{\mu\nu} = \partial_\mu A_\nu-\partial_\nu A_\mu$ is the gauge-field strength tensor, $V(\phi)$ is the inflaton potential, $f(\phi)$ is the kinetic coupling function, $g_{\mu\nu}$ represents the FLRW metric in Eq.~\eqref{eq:FLRW-metric}, and $g=\operatorname{det}g_{\mu\nu}$ is its determinant.

Varying the action with respect to the metric yields the energy--momentum tensor,
\begin{multline}
\label{EMT-general}
    T_{\mu\nu}
    %=\frac{2}{\sqrt{-g}}\frac{\delta S}{\delta g^{\mu\nu}}
    =\frac{f^2(\phi)}{4}
g_{\mu\nu}F_{\alpha\beta}F^{\alpha\beta}
-f^2(\phi)g^{\rho\kappa}F_{\mu\rho}F_{\nu\kappa}\\
+\partial_\mu\phi\partial_\nu\phi-g_{\mu\nu}\left(\frac{1}{2}g^{\rho\kappa}\partial_\rho\phi\partial_\kappa\phi-V(\phi)\right)\,.
\end{multline}

The equation of motion for the inflaton field is obtained by varying the action with respect to $\phi$:
\begin{multline}
\label{rawFieldEq}
\frac{1}{\sqrt{-g}}
\partial_\mu
\left(
\sqrt{-g}\,g^{\mu\nu}\partial_\nu\phi
\right)
+\frac{\mathrm{d}V(\phi)}{\mathrm{d}\phi}
\\
=
-\frac{1}{2}
f(\phi)\frac{\mathrm{d} f(\phi)}{\mathrm{d}\phi}
F_{\mu\nu}F^{\mu\nu}.
\end{multline}

The evolution of the gauge field is governed by the Maxwell equations and the Bianchi identities,
\begin{equation}
\label{rawMaxwellEq}
\frac{1}{\sqrt{-g}}
\partial_\mu
\left(
\sqrt{-g}\,f^2(\phi)F^{\mu\nu}
\right)
=0,
\end{equation}
and
\begin{equation}
\label{rawBianchiEq}
\frac{1}{\sqrt{-g}}
\partial_\mu
\left(
\sqrt{-g}\,\widetilde{F}^{\mu\nu}
\right)
=0,
\end{equation}
respectively. Here,
\begin{equation}
\widetilde{F}^{\mu\nu}
=
\frac{1}{2\sqrt{-g}}
\epsilon^{\mu\nu\alpha\beta}F_{\alpha\beta}
\end{equation}
is the dual field-strength tensor, where $\epsilon^{\mu\nu\alpha\beta}$ is the four-dimensional Levi-Civita symbol with $\epsilon^{0123}=1$.

The components of the field-strength tensor can be expressed in terms of the electric and magnetic fields defined in the orthonormal physical basis~\cite{landau} as
\begin{equation}
\label{magnAndElectrickFields}
F_{0i}=aE_{(i)},
\qquad
F^{0i}=-\frac{1}{a}E_{(i)},
\end{equation}
and
\begin{equation}
F_{ij}=-a^2\epsilon_{ijk}B_{(k)},
\qquad
F^{ij}=-\frac{1}{a^2}\epsilon^{ijk}B_{(k)}.
\end{equation}
Here, $(i)$ denotes components in the orthonormal physical basis, while $\epsilon_{ijk}$ is the three-dimensional Levi-Civita symbol. 

We now specialize to a spatially homogeneous inflaton background, $\phi=\phi(t)$. Consequently, the kinetic coupling function is homogeneous as well, $f(\phi)=f(t)$, and the gauge-field equations are invariant under spatial translations. In this setting, we can choose the Coulomb gauge for the gauge field, $A_\mu = (0,-\boldsymbol{A})$ with $\operatorname{div}\boldsymbol{A}=0$. With this choice, the electric and magnetic fields measured in the orthonormal physical basis can be expressed directly in terms of the spatial vector potential as
\begin{equation}
\label{E_and_B}
\boldsymbol{E}
=-\frac{1}{a}\dot{\boldsymbol{A}},
\qquad
\boldsymbol{B}
=\frac{1}{a^2}\,\mathrm{rot}\,\boldsymbol{A}.
\end{equation}
In terms of the physical electric and magnetic fields, Eqs.~\eqref{rawMaxwellEq}--\eqref{rawBianchiEq} take the form
\begin{equation}
\label{Maxwell}
\dot{\boldsymbol{E}}
+2H\boldsymbol{E}
+2\frac{\dot f}{f}\boldsymbol{E}
-\frac{1}{a}\mathrm{rot}\,\boldsymbol{B}
=0,
\end{equation}
together with
\begin{equation}
\dot{\boldsymbol{B}}
+2H\boldsymbol{B}
+\frac{1}{a}\mathrm{rot}\,\boldsymbol{E}
=0,
\end{equation}
and the constraints
\begin{equation}
\mathrm{div}\,\boldsymbol{E}=0,
\qquad
\mathrm{div}\,\boldsymbol{B}=0.
\end{equation}

During inflation, we treat the gauge field as a quantum field interacting with the classical background. Its contribution to the background dynamics is characterized by the vacuum expectation values of quadratic combinations of the electric and magnetic field operators. In particular, the gauge-field energy density is obtained by taking the vacuum expectation value of the gauge-field contribution to the \(00\) component of the energy--momentum tensor in Eq.~\eqref{EMT-general}:
\begin{equation}
    \rho_{\mathrm{GF}}
    = \langle T_{00}^{\mathrm{GF}}\rangle
    = \rho_E+\rho_B\,,
\end{equation}
where we have decomposed it into electric and magnetic contributions,
\begin{equation}
\label{rhoECorrelator}
\rho_E
=
\frac{f^2}{2}
\langle 0|\hat{\boldsymbol{E}}^2|0\rangle,
\end{equation}
and
\begin{equation}
\label{rhoBCorrelator}
\rho_B
=
\frac{f^2}{2}
\langle 0|\hat{\boldsymbol{B}}^2|0\rangle.
\end{equation}

For a homogeneous inflaton field, the Klein--Gordon equation \eqref{rawFieldEq} becomes
\begin{equation}
\label{inflatonEq}
\ddot{\phi}
+3H\dot{\phi}
+\frac{\mathrm{d}V(\phi)}{\mathrm{d}\phi}
=
2\frac{1}{f(\phi)}
\frac{\mathrm{d}f(\phi)}{\mathrm{d}\phi}
\left(\rho_E-\rho_B\right).
\end{equation}
Under the assumption of a homogeneous and isotropic background, the expectation values $\rho_E$ and $\rho_B$ depend only on time.

Finally, the first Friedmann equation takes the form
\begin{equation}
\label{Friedman}
H^2
=
\frac{1}{3M_{\mathrm{P}}^2}
\left(
\frac{\dot{\phi}^2}{2}
+V(\phi)
+\rho_{\mathrm{GF}}
\right),
\end{equation}
where the first two terms in parentheses describe the energy density of the homogeneous inflaton field.

\subsection{Gauge-field dynamics}
\label{subsec:gauge_field_dynamcics}

The nonstandard kinetic term in the action makes the gauge field $A_\mu$ noncanonical. It is therefore convenient to introduce the canonically normalized field
\begin{equation}
\label{cannonicalField}
\mathcal{A}_\mu=f(\phi)A_\mu.
\end{equation}
Since the Coulomb gauge has already been imposed in the previous subsection, we will use the same notation $\boldsymbol{\mathcal{A}}$ for the corresponding canonical spatial three-vector.

It is convenient to decompose the transverse field $\boldsymbol{\mathcal{A}}$ in Fourier space into circular polarization states, described by the polarization vectors $\boldsymbol{\epsilon}^{\lambda}(\boldsymbol{k})$, $\lambda=\pm1$, which are orthogonal to the direction of propagation defined by the comoving momentum $\boldsymbol{k}$. Their properties are
\begin{equation}
\boldsymbol{k}\cdot\boldsymbol{\epsilon}^{\lambda}(\boldsymbol{k})=0,\qquad i\left[\boldsymbol{k}\times\boldsymbol{\epsilon}^{\lambda}(\boldsymbol{k})\right]
=\lambda k\boldsymbol{\epsilon}^{\lambda}(\boldsymbol{k}),
\end{equation}
\begin{equation}
\boldsymbol{\epsilon}^{\lambda}(\boldsymbol{k})
\cdot
\boldsymbol{\epsilon}^{*\lambda'}(\boldsymbol{k})
=\delta^{\lambda\lambda'},
\quad
\sum_{\lambda=\pm1}
\epsilon_i^\lambda(\boldsymbol{k})
\epsilon_j^{*\lambda}(\boldsymbol{k})
=
\delta_{ij}-\frac{k_i k_j}{k^2},
\end{equation}
where $\delta_{ij}$ is the Kronecker delta.

For convenience, we now switch to conformal time $\eta$. The canonical field operator can then be expanded in Fourier modes as
\begin{multline}
\label{operator}
\hat{\mathcal{A}}_i(\eta,\boldsymbol{x})
=
\sum_{\lambda=\pm 1}
\int
\frac{\mathrm{d}^3 \boldsymbol{k}}
{(2\pi)^{3/2}\sqrt{2k}}\\
\times\bigg[
\epsilon_i^\lambda(\boldsymbol{k})
e^{i\boldsymbol{k}\cdot\boldsymbol{x}}
\mathcal{A}_\lambda(\eta,k)
\hat a_{\boldsymbol{k}\lambda} + \text{H.c.}\bigg],
\end{multline}
where ``H.c.'' denotes the Hermitian conjugate term; $\hat a_{\boldsymbol{k}\lambda}$ and $\hat a_{\boldsymbol{k}\lambda}^\dagger$ are the annihilation and creation operators, respectively, satisfying the standard commutation relations
\begin{equation}
\label{commut}
    [\hat a_{\boldsymbol{k}\lambda},
    \hat a_{\boldsymbol{k}'\lambda'}^\dagger]
    =
    \delta_{\lambda\lambda'}
    \delta^{(3)}(\boldsymbol{k}-\boldsymbol{k}').
\end{equation}
The Bunch-Davies vacuum satisfies
\begin{equation}
\label{AnnOp}
    \hat a_{\boldsymbol{k}\lambda}|0\rangle=0.
\end{equation}

The Fourier mode function $\mathcal{A}_\lambda$ satisfies the equation
\begin{equation}
\label{eqBRaw}
    \mathcal{A}_\lambda''
    +
    \left(
        k^2-\frac{f''}{f}
    \right)
    \mathcal{A}_\lambda
    =0.
\end{equation}
We refer to $\mathcal{A}_\lambda$ as the \textit{magnetic mode function}. It is normalized by the Bunch--Davies initial condition
\begin{equation}
\label{BDCondition}
    \mathcal{A}_\lambda
    \underset{k|\eta|\to\infty}{\longrightarrow}
    e^{-ik\eta}.
\end{equation}
For convenience, we introduce an auxiliary mode function associated with the electric field,
\begin{equation}
\label{definitionD}
    \mathcal{D}_\lambda
    =
    \frac{f}{k}
    \frac{\partial}{\partial\eta}
    \left(
        \frac{\mathcal{A}_\lambda}{f}
    \right).
\end{equation}
We refer to $\mathcal{D}_\lambda$ as the \textit{electric mode function}. It satisfies
\begin{equation}
\label{eqDRaw}
    \mathcal{D}_\lambda''
    +
    \left[
        k^2
        +\frac{f''}{f}
        -2\left(\frac{f'}{f}\right)^2
    \right]
    \mathcal{D}_\lambda
    =0,
\end{equation}
and is normalized by the condition 
\begin{equation}
\label{BDConditionElectricMode}
    \mathcal{D}_\lambda
    \underset{k|\eta|\to\infty}{\longrightarrow}
    -ie^{-ik\eta}.
\end{equation}

Since the equations for the two polarizations are identical and modes with different polarizations do not interact, we suppress the polarization index below and write simply $\mathcal{A}$ and $\mathcal{D}$.

Using Eqs.~\eqref{magnAndElectrickFields}, \eqref{cannonicalField}, and \eqref{operator}, the vacuum correlators defined in Eqs.~\eqref{rhoECorrelator} and \eqref{rhoBCorrelator} can be expressed in terms of the mode functions as
\begin{equation}
\label{rhoERaw}
    \rho_E
    =
    \frac{1}{a^4}\int
    \frac{\mathrm{d}^3\boldsymbol{k}}{(2\pi)^3}\,\frac{k}{2}
    \left|\mathcal{D}\right|^2,
\end{equation}
and
\begin{equation}
\label{rhoBRaw}
    \rho_B
    =
    \frac{1}{a^4}\int
    \frac{\mathrm{d}^3\boldsymbol{k}}{(2\pi)^3}\,\frac{k}{2}
    \left|\mathcal{A}\right|^2.
\end{equation}

In Fourier space, the magnetic mode function $\mathcal{A}$ and the electric mode function $\mathcal{D}$ form a pair of dual variables in the Hamiltonian formulation. Their evolution can be written as the first-order system
\begin{equation}
\label{hamiltonEq}
    \frac{\partial}{\partial\eta}
    \begin{pmatrix}
        \mathcal{A}\\
        \mathcal{D}
    \end{pmatrix}
    =
    \begin{pmatrix}
        {f'}/{f} & k\\[4pt]
        -k & -{f'}/{f}
    \end{pmatrix}
    \begin{pmatrix}
        \mathcal{A}\\
        \mathcal{D}
    \end{pmatrix}.
\end{equation}
This representation makes manifest a symmetry of the equations under reversal of the sign of $f'/f$. Indeed, at the level of the dynamical
system, the transformation 
\begin{equation}
\label{symmetryIncludingPhase}
    \frac{f'}{f}\rightarrow-\frac{f'}{f}
    \qquad\Longrightarrow\qquad
    \mathcal{A}\rightarrow i\mathcal{D},
    \qquad
    \mathcal{D}\rightarrow-i\mathcal{A}
\end{equation}
maps solutions of Eq.~\eqref{hamiltonEq} into solutions.
The phase factors $i$ and $-i$ are chosen in a way that the transformation remains consistent with the boundary conditions \eqref{BDCondition} and \eqref{BDConditionElectricMode}. 
Since the energy densities are proportional to $|\mathcal{A}|^2$ and $|\mathcal{D}|^2$, respectively, and the overall phase does not affect the absolute values of the mode functions, this transformation implies
\begin{equation}
\label{symmetry}
    \frac{f'}{f}\rightarrow-\frac{f'}{f}
    \qquad\Longrightarrow\qquad
    \rho_E\rightarrow\rho_B,
    \qquad
    \rho_B\rightarrow\rho_E,
\end{equation}
while preserving the total gauge-field energy density $\rho_{\mathrm{GF}}$.%
\footnote{This symmetry is closely related to the discrete electric--magnetic duality transformation of source-free Maxwell theory, corresponding to a $\pi/2$ rotation of the continuous $SO(2)$ duality group, under which $\boldsymbol{E}\to\boldsymbol{B}$ and $\boldsymbol{B}\to-\boldsymbol{E}$ while the electromagnetic energy--momentum tensor remains invariant~\cite{Calkin:1965,Deser:1976iy,Deser:1981fr}. In the present model, however, the transformation also reverses the logarithmic derivative of the gauge kinetic coupling, corresponding to $f\to f^{-1}$. It is therefore more precisely interpreted as a duality map between models with reciprocal kinetic couplings, which exchanges the electric and magnetic sectors. Related duality transformations in inflationary gauge-field models have been discussed in Ref.~\cite{Giovannini:2013rza}.}

The expressions \eqref{rhoERaw} and \eqref{rhoBRaw} provide the gauge-field energy densities as integrals over all comoving momenta. Since the mode functions approach their vacuum form at large momenta,
\begin{equation}
\mathcal{A},\,\mathcal{D}
\underset{k\to\infty}{\sim}
e^{-ik\eta},
\end{equation}
the integrals in Eqs.~\eqref{rhoERaw} and \eqref{rhoBRaw} diverge as $k^4$ in the UV. Consequently, both energy densities contain a UV-divergent vacuum contribution. The total gauge-field energy density, $\rho_{\mathrm{GF}}$,
is therefore not directly defined by the unregulated expressions. 
This UV divergence reflects the contribution of quantum fluctuations with arbitrarily large physical momenta and must therefore be treated by an appropriate regularization prescription. 

In the following sections, we consider two different approaches to this problem. We first examine conventional UV regularization using a physical-momentum cutoff and dimensional regularization in Sec.~\ref{sec:ultraviolet_regularization}, and then compare these results with horizon-based prescriptions in Sec.~\ref{sec:approach_comparison}.

\section{Conventional QFT regularization}
\label{sec:ultraviolet_regularization}

\subsection{Analytical mode functions}
\label{subsec:analytic_mode_solutions}

To evaluate the gauge-field energy density, we first derive the analytical solutions for the mode functions during inflation. We work at zeroth order in the slow-roll parameters, so that the inflaton velocity and the Hubble rate can be treated as approximately constant,
\begin{equation}
\label{approx}
    \dot{\phi}\simeq \mathrm{const.},
    \qquad
    H\simeq \mathrm{const.},
    \qquad
    aH=-\frac{1}{\eta}.
\end{equation}
We further assume that the logarithmic rate of change of the kinetic coupling is approximately constant and introduce the parameter
\begin{equation}
    \gamma_1
    \equiv
    \frac{\dot f}{fH}=\text{const}.
\end{equation}
It then follows that
\begin{equation}
    f=a^{\gamma_1},
\end{equation}
and
\begin{equation}
    \frac{\ddot f}{fH^2}
    \equiv \gamma_2
    \simeq \mathrm{const.},
    \qquad
    \gamma_2=\gamma_1^2.
\end{equation}

Under these assumptions, the equations for the magnetic and electric mode functions, Eqs.~\eqref{eqBRaw} and \eqref{eqDRaw}, become
\begin{equation}
\label{eqBSim}
    \mathcal{A}''
    +
    \left[
        k^2
        -
        \frac{\gamma_1(\gamma_1+1)}{\eta^2}
    \right]
    \mathcal{A}
    =0,
\end{equation}
and
\begin{equation}
\label{eqDSim}
    \mathcal{D}''
    +
    \left[
        k^2
        -
        \frac{\gamma_1(\gamma_1-1)}{\eta^2}
    \right]
    \mathcal{D}
    =0.
\end{equation}

Introducing the dimensionless variable
\begin{equation}
\label{xdef}
    x=-k\eta=\frac{k}{aH},
\end{equation}
the equation for the magnetic mode function is of Bessel type. The solution satisfying the Bunch--Davies condition \eqref{BDCondition} is
\begin{equation}
\label{solutionA}
    \mathcal{A}
    =
    \sqrt{\frac{\pi}{2}}\,
    e^{i\left(\frac{\pi}{4}+\frac{\pi}{2}\alpha\right)}
    \sqrt{x}\,
    H^{(1)}_{\alpha}(x),
    \qquad
    \alpha=
    \left|\gamma_1+\frac{1}{2}\right|.
\end{equation}

The electric mode function can be obtained independently from Eq.~\eqref{eqDSim}. However, the two mode equations are related by a symmetry of the first-order system \eqref{hamiltonEq}. Under the transformation
\begin{equation}
\label{fInverseSymmetry}
    f\rightarrow\frac{1}{f},
\end{equation}
we have
\begin{equation}
    \frac{f'}{f}\rightarrow-\frac{f'}{f},
    \qquad
    \gamma_1\rightarrow-\gamma_1,
\end{equation}
while the magnetic and electric mode functions are interchanged according to \eqref{symmetryIncludingPhase}.
Consequently, provided that the regularization and subtraction schemes preserve this symmetry, the corresponding energy densities satisfy
\begin{equation}
\label{rhoSymmetryGamma}
    \rho_E(\gamma_1)
    =
    \rho_B(-\gamma_1).
\end{equation}

This relation allows us to determine the electric energy density entirely from the magnetic one. In particular, it is sufficient to evaluate the magnetic contribution for arbitrary $\gamma_1$ and obtain the electric contribution by the replacement $\gamma_1 \to -\gamma_1$. We therefore restrict the calculations in the following subsections to the magnetic energy density.

The magnetic energy density in Eq.~\eqref{rhoBRaw} can equivalently be expressed in terms of the physical momentum $p=k/a$ as
\begin{equation}
\label{rhoBPhysicalMomentum}
    \rho_B
    =
    \int
    \frac{\mathrm{d}^3\boldsymbol{p}}{(2\pi)^3}
    \frac{p}{2}
    \left|\mathcal{A}\right|^2.
\end{equation}
Substituting the analytical solution \eqref{solutionA} then gives
\begin{equation}
\label{rhoBPhysicalMomentumHankel}
    \rho_B
    =
    \frac{\pi}{4H}\int
    \frac{\mathrm{d}^3\boldsymbol{p}}{(2\pi)^3}
    p^2
    \left|
        H^{(1)}_{\alpha}\left(\frac{p}{H}\right)
    \right|^2,
    \qquad
    \alpha=
    \left|\gamma_1+\frac{1}{2}\right|.
\end{equation}

The integral in Eq.~\eqref{rhoBPhysicalMomentumHankel} is UV divergent and therefore requires regularization. It may also exhibit an infrared divergence for sufficiently large $|\gamma_1|$. From the small-$x$ behavior of the Hankel function, the magnetic contribution is infrared finite for $-3<\gamma_1<2$. The symmetry \eqref{rhoSymmetryGamma} implies that the electric contribution is finite in the infrared for $-2<\gamma_1<3$. We therefore restrict the analysis to
\begin{equation}
    |\gamma_1|<2,
\end{equation}
which guarantees infrared convergence of both the electric and magnetic contributions.

\subsection{Hard-cutoff regularization}
\label{subsec:physical_cutoff}

The energy density in Eq.~\eqref{rhoBPhysicalMomentumHankel} contains contributions from modes with arbitrarily large physical momenta and is therefore UV divergent. To regulate the UV sector, we introduce a physical momentum cutoff $\Lambda$ through the condition
\begin{equation}
|\boldsymbol{p}|\leq\Lambda\, .
\end{equation}

The regulated magnetic energy density is consequently defined as
\begin{align}
\label{rhoBLambda}
\rho_B^{(\Lambda)}
&=
\frac{\pi}{4H}\int \frac{\mathrm{d}\Omega}{(2\pi)^3}
\int_0^{\Lambda}
\mathrm{d}p\,p^4
\left|
        H^{(1)}_{\alpha}\left(\frac{p}{H}\right)
    \right|^2 \nonumber\\
&= \frac{H^4}{8\pi}\int_0^{\Lambda/H}
\mathrm{d}x\,x^4
\left|
        H^{(1)}_{\alpha}\left(x\right)
    \right|^2.
\end{align}
The corresponding total gauge-field energy density is
\begin{equation}
\label{rhoGFLambda}
\rho_{\mathrm{GF}}^{(\Lambda)}(\gamma_1)
=
\rho_B^{(\Lambda)}(\gamma_1)+\rho_B^{(\Lambda)}(-\gamma_1).
\end{equation}

The cutoff $\Lambda$ regulates the UV sector. With the finite upper integration limit, the integral in Eq.~\eqref{rhoBLambda} can be evaluated in terms of generalized hypergeometric functions, as shown in Appendix~\ref{app:Hankel-integrals}; see Eq.~\eqref{integral-Hankel-finite-upper-limit}. The resulting expression still contains divergent contributions as $\Lambda\rightarrow\infty$. In this limit, the integral can be expanded asymptotically. The resulting expressions contain quartic, quadratic, and logarithmic UV divergences. Using the asymptotic expansion given in Appendix~\ref{app:Hankel-integrals}, Eq.~\eqref{int-Hankel-finite-upper-limit-asympt}, we find
\begin{multline}
\label{rhoBGammaRenorm}
\rho_B^{(\Lambda)}
=\frac{\Lambda^4}{16\pi^2}
+
\frac{\gamma_1(\gamma_1+1)H^2\Lambda^2}{16\pi^2}\\
+
\frac{3\gamma_1(\gamma_1^2-1)(\gamma_1+2)H^4}{32\pi^2}
\ln\left(\frac{\Lambda}{H}\right)
\\
+
\frac{\gamma_1(\gamma_1^2-1)(\gamma_1+2)H^4}{128\pi^2}
\bigg[-7+6\ln 4\\
-6\psi^{(0)}(\gamma_1+2)
-6\psi^{(0)}(-\gamma_1+2) -\frac{6}{\gamma_1+2}
\bigg].
\end{multline}
Here, $\psi^{(0)}(x)=\mathrm{d}\ln\Gamma(x)/\mathrm{d}x$ denotes the digamma function, while $\Gamma(x)$ is the gamma function.

Using the symmetry relation \eqref{rhoSymmetryGamma}, the total gauge-field energy density is obtained as
\begin{multline}
\label{rhoGFhard-cutoff-expansion}
\rho_{\mathrm{GF}}^{(\Lambda)}
=
\frac{\Lambda^4}{8\pi^2}
+
\frac{\gamma_1^2H^2\Lambda^2}{8\pi^2}
+
\frac{3\gamma_1^2(\gamma_1^2-1)H^4}{32\pi^2}\bigg[2
\ln\left(\frac{\Lambda}{H}\right)\\
+\ln 4-\frac{7}{6}
-\psi^{(0)}(\gamma_1+2)
-\psi^{(0)}(-\gamma_1+2)
\bigg].
\end{multline}
In obtaining the explicitly even form in $\gamma_1$, we used the recurrence relation for the digamma function, $\psi^{(0)}(1+x)=\psi^{(0)}(x)+1/x$.

The first three terms in these expressions contain the UV divergences, while the remaining terms are finite as $\Lambda\rightarrow\infty$. In Sec.~\ref{subsec:dimensional_regularization}, we obtain the corresponding result using dimensional regularization and use it to cross-check the UV structure found with the hard cutoff. The prescription for fixing the remaining finite part is discussed in Sec.~\ref{subsec:regularized-en-dens}.

\subsection{Dimensional regularization}
\label{subsec:dimensional_regularization}

As a cross-check of the UV structure obtained with the hard cutoff, we now evaluate the gauge-field energy density using dimensional regularization. We analytically continue the spatial momentum integral from three to $d-1$ dimensions, with
\begin{equation}
d=4-\epsilon,
\end{equation}
where $\epsilon$ is the dimensional regulator. The physical momentum measure is correspondingly replaced according to
\begin{equation}
\int \frac{\mathrm{d}^3\boldsymbol{p}}{(2\pi)^3}
\longrightarrow
\mu^{4-d}
\int \frac{\mathrm{d}^{d-1}\boldsymbol{p}}{(2\pi)^{d-1}},
\end{equation}
where $\mu$ is the renormalization scale introduced to keep the energy density dimensionally consistent. With this replacement, the regularized expression for the magnetic energy density follows directly from Eq.~\eqref{rhoBPhysicalMomentumHankel} and reads as
\begin{align}
    \rho_B^{\mathrm{DR}}
    &=
    \frac{\pi \mu^{4-d}}{4H}\int
    \frac{\mathrm{d}^{d-1}\boldsymbol{p}}{(2\pi)^{d-1}}
    p^2
    \left|
        H^{(1)}_{\alpha}\left(\frac{p}{H}\right)
    \right|^2\nonumber\\
    &=\frac{\pi \mu^{4-d}}{4H}\int \frac{\mathrm{d}\Omega_{d-2}}{(2\pi)^{d-1}}\int_0^{+\infty}\mathrm{d}p\,p^{d}\,\left|
        H^{(1)}_{\alpha}\left(\frac{p}{H}\right)
    \right|^2\nonumber\\
    &=H^4\frac{2^{-d} \pi^{\frac{3}{2}-\frac{d}{2}}}{\Gamma \left(\frac{d-1}{2}\right)}\Big(\frac{\mu}{H}\Big)^{4-d} \int_0^{+\infty}\mathrm{d}x\,x^{d}\,\left|
        H^{(1)}_{\alpha}\left(x\right)
    \right|^2\, ,
    \label{rhoBdimreg-integral}
\end{align}
where in the last step we introduced the dimensionless variable $x=p/H$ and used the explicit expression for the total solid angle on a $(d-2)$-dimensional unit sphere
\begin{equation}
    \int \mathrm{d}\Omega_{d-2} = \frac{2 \pi ^{\frac{d-1}{2}}}{\Gamma \left(\frac{d-1}{2}\right)}\, .
\end{equation}
The integral over $x$ in Eq.~\eqref{rhoBdimreg-integral} can be computed in terms of the gamma functions and the full result is shown in Appendix~\ref{app:Hankel-integrals}; see Eq.~\eqref{app-integral-dimreg-full}. 

The resulting expression is expanded around $d=4$, or equivalently $\epsilon=0$, to isolate the divergent and finite contributions. Substituting $\alpha = \left|\gamma_1+\frac{1}{2}\right|$ into Eq.~\eqref{app-integral-dimreg-expansion}, we obtain, up to terms that vanish in the limit $\epsilon\to 0$,
\begin{multline}
    \rho_B^{\mathrm{DR}}(\gamma_1)=\frac{3 \gamma_1 \left(\gamma_1^2-1\right) \left(\gamma_1+2\right)H^4}{64 \pi ^2} \bigg[\frac{2}{\epsilon}+\ln(4\pi)-\gamma_E+\frac{5}{6}\\
    +\ln\frac{\mu^2}{H^2}
    -\psi ^{(0)}\left(\gamma_1+2\right)-\psi ^{(0)}\left(-\gamma_1+2\right)-\frac{1}{\gamma_1+2}\bigg],
\label{rhoBdimreg-expansion}
\end{multline}
where $\gamma_E$ denotes the Euler--Mascheroni constant.

The magnetic contribution can again be used to obtain the electric contribution through the symmetry relation \eqref{rhoSymmetryGamma}. Thus, the total gauge-field energy density is obtained as
\begin{align}
\rho_{\mathrm{GF}}^{\mathrm{DR}}
&=
\rho_B^{\mathrm{DR}}(\gamma_1)
+
\rho_B^{\mathrm{DR}}(-\gamma_1)\nonumber\\
&=\frac{3 \gamma_1^2 \left(\gamma_1^2-1\right)H^4}{32 \pi ^2} \bigg[2\bigg(\frac{1}{\epsilon}+\ln\frac{\mu}{H}\bigg)-\gamma_E+\frac{5}{6}\nonumber\\
    &
    +\ln(4\pi)-\psi ^{(0)}\left(\gamma_1+2\right)-\psi ^{(0)}\left(-\gamma_1+2\right)\bigg].
\label{rhoGFdimreg-expansion}
\end{align}

In dimensional regularization, the power divergences that appear with the hard cutoff do not arise explicitly; instead, the UV divergence is encoded in the pole in $\epsilon$. The agreement of the coefficient of this pole with the corresponding logarithmically divergent term in the hard-cutoff regularization provides a consistency check of the UV structure obtained above. 

The remaining finite terms depend on the regularization and subtraction prescription and will be considered together in Sec.~\ref{subsec:regularized-en-dens}.

\subsection{Finite part of the energy density}
\label{subsec:regularized-en-dens}

In a complete covariant renormalization of the theory, the UV divergences of the gauge-field energy--momentum tensor are absorbed into local counterterms involving the gravitational and matter sectors. However, the calculation presented here is performed after imposing the zeroth-order slow-roll approximation \eqref{approx}, in which $H$ and $\dot{\phi}$ are treated as approximately constant and the time dependence of the kinetic coupling is characterized by the single parameter $\gamma_1$. The resulting reduced system does not retain the full set of independent background structures required to identify the corresponding local counterterms and to fix their finite parts unambiguously. We therefore regard the expressions obtained above as regulator-dependent representations of the UV structure and fix the remaining finite contribution through a separate finite-part prescription.

The finite parts of the expressions computed in the two regularization schemes are not uniquely fixed by the UV divergences alone. In particular, a rescaling of the cutoff $\Lambda$ in hard-cutoff regularization or of the renormalization scale $\mu$ in dimensional regularization changes the finite contribution associated with the logarithmic divergence. We therefore parameterize the remaining finite-part ambiguity by a single dimensionless constant $C$.

After subtracting the divergent terms, the regular part of the total gauge-field energy density can be written in the common form
\begin{equation}
\label{rhoGFreg-general}
\rho_{\mathrm{GF}}^{\mathrm{reg}}
=
\frac{3\gamma_1^2(\gamma_1^2-1)H^4}{32\pi^2}
\bigg[
C-\psi^{(0)}\left(\gamma_1+2\right)
-\psi^{(0)}\left(-\gamma_1+2\right)
\bigg],
\end{equation}
where $C$ is an undetermined constant.

We fix this constant by \textit{requiring} the regularized, coupling-induced energy density to be nonnegative over the range of $\gamma_1$ considered here. The overall prefactor in Eq.~\eqref{rhoGFreg-general} vanishes at $\gamma_1=\pm1$, so these points are roots of the regularized energy density independently of $C$. The finite-part constant determines the behavior of the energy density in their vicinity and, consequently, the position of the global minimum. Without loss of generality, we expand Eq.~\eqref{rhoGFreg-general} around $\gamma_1=1$ and obtain
\begin{equation}
    \rho_{\mathrm{GF}}^{\mathrm{reg}}
    \underset{\gamma_1\to 1}{=}
    \frac{3H^4}{32\pi^2}
    (1-\gamma_1)
    \left(3-4\gamma_E-2C\right)
    +\mathcal{O}\left((\gamma_1-1)^2\right).
\end{equation}
Requiring the global minimum to coincide with the roots at $\gamma_1=\pm1$ fixes the finite-part constant to
\begin{equation}
\label{Cvalue}
    C=\frac{3}{2}-2\gamma_E.
\end{equation}

With the finite-part prescription specified above, the regularized total gauge-field energy density is
\begin{multline}
\label{rhoGFreg-final}
\rho_{\mathrm{GF}}^{\mathrm{reg}}(\gamma_1)
=
\frac{3\gamma_1^2(\gamma_1^2-1)H^4}{32\pi^2}
\bigg[\frac{3}{2}-2\gamma_E\\
-\psi^{(0)}\left(\gamma_1+2\right)
-\psi^{(0)}\left(-\gamma_1+2\right)
\bigg].
\end{multline}

The corresponding magnetic energy density reads as
\begin{multline}
\label{rhoBreg-final}
\rho_{B}^{\mathrm{reg}}(\gamma_1)
=
\frac{3\gamma_1(\gamma_1^2-1)(\gamma_1+2)H^4}{64\pi^2}
\bigg[\frac{3}{2}-2\gamma_E-\frac{1}{\gamma_1+2}\\
-\psi^{(0)}\left(\gamma_1+2\right)
-\psi^{(0)}\left(-\gamma_1+2\right)
\bigg],
\end{multline}
while the electric contribution follows from $\rho_{E}^{\mathrm{reg}}(\gamma_1)=\rho_{B}^{\mathrm{reg}}(-\gamma_1)$.

\begin{figure}[t]
  \centering
  \includegraphics[width=0.99\columnwidth]{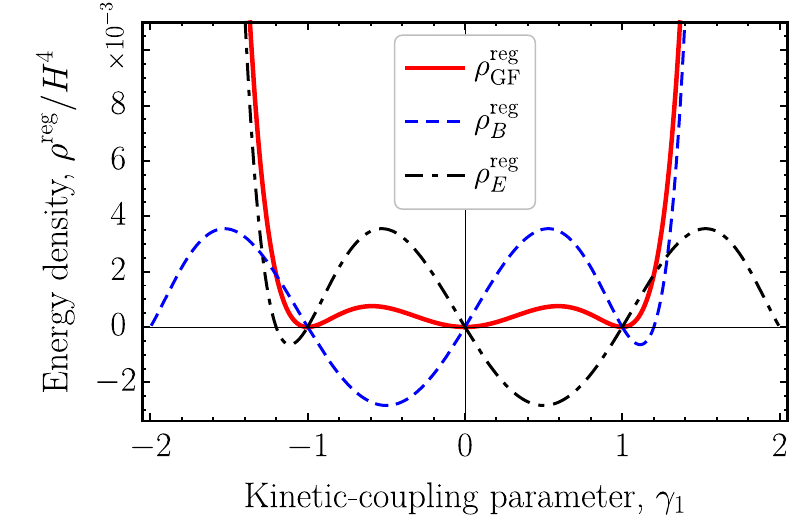}
  \caption{Regularized gauge-field energy densities, in Hubble units, as functions of the kinetic-coupling parameter $\gamma_1$. The total energy density $\rho_{\mathrm{GF}}^{\mathrm{reg}}$ is shown by the red solid curve, while the magnetic and electric contributions, $\rho_B^{\mathrm{reg}}$ and $\rho_E^{\mathrm{reg}}$, are shown by the blue dashed and black dashed--dotted curves, respectively. The total energy density is nonnegative and vanishes at $\gamma_1=0,\,\pm 1$, whereas the individual electric and magnetic contributions can become negative over parts of the parameter range shown.}
  \label{fig:rho_vec_E_B_reg}
\end{figure}

Figure~\ref{fig:rho_vec_E_B_reg} shows the regularized magnetic and electric energy densities (in Hubble units) by the blue dashed and black dashed--dotted lines, respectively, and the total gauge-field energy density by the red solid line as functions of $\gamma_1$. By construction, the total energy density is nonnegative over the range of $\gamma_1$ considered here and vanishes at $\gamma_1=\pm1$. It also vanishes at $\gamma_1=0$, corresponding to a constant kinetic coupling and hence to the absence of the coupling-induced contribution to the regularized energy density.

The special values $\gamma_1=\pm1$ have a direct interpretation in terms of the mode equations. At $\gamma_1=-1$, the coupling-dependent term in the magnetic mode equation \eqref{eqBSim} vanishes, while at $\gamma_1=1$ the corresponding term in the electric mode equation \eqref{eqDSim} vanishes. In these cases, the respective mode exhibits vacuum-like behavior.

The total gauge-field energy density is nonnegative over the range of $\gamma_1$ considered here, as imposed by the finite-part prescription. In contrast, the individual electric and magnetic contributions can become negative for some values of $\gamma_1$. In fact, no single choice of the finite-part constant $C$ can render both $\rho_E^{\mathrm{reg}}$ and $\rho_B^{\mathrm{reg}}$ nonnegative over the entire range of $\gamma_1$. This indicates that, after regularization, the electric and magnetic contributions need not be separately nonnegative. We therefore focus on the total gauge-field energy density in the following discussion.

For completeness, the subtraction term in hard-cutoff regularization is
\begin{multline}
\label{rhoGFhard-div}
\rho_{\mathrm{GF}}^{\mathrm{sub},\Lambda}
=
\frac{\Lambda^4}{8\pi^2}
+
\frac{\gamma_1^2H^2\Lambda^2}{8\pi^2}
\\+
\frac{3\gamma_1^2(\gamma_1^2-1)H^4}{16\pi^2}\bigg[
\ln\left(\frac{\Lambda}{H}\right)
+\ln 2-\frac{4}{3}+\gamma_E \bigg].
\end{multline}
% Thus,
% \begin{equation}
% \rho_{\mathrm{GF}}^{\mathrm{reg}}
% =
% \lim_{\Lambda\to\infty}
% \left[
% \rho_{\mathrm{GF}}^{(\Lambda)}
% -
% \rho_{\mathrm{GF}}^{\mathrm{sub},\Lambda}
% \right],
% \end{equation}
% with the finite-part prescription fixed by Eq.~\eqref{Cvalue}.

In dimensional regularization, the subtraction term can be written as
\begin{align}
\label{rhoGFdimreg-div}
\rho_{\mathrm{GF}}^{\mathrm{sub},\mathrm{DR}}
&=
\frac{3 \gamma_1^2 \left(\gamma_1^2-1\right)H^4}{32 \pi ^2}
\bigg[
\frac{2}{\epsilon}-\gamma_E+\ln(4\pi)
\bigg]\nonumber\\
&\quad+
\frac{3 \gamma_1^2 \left(\gamma_1^2-1\right)H^4}{16 \pi ^2}
\bigg[
\ln\frac{\mu}{H}+\gamma_E-\frac{1}{3}
\bigg].
\end{align}
The first line corresponds to the standard $\overline{\mathrm{MS}}$ subtraction, while the second line ensures that the resulting finite part agrees with the prescription obtained from the hard-cutoff regularization. Equivalently, since the second line vanishes for $\mu=H e^{\frac{1}{3}-\gamma_E}\approx 0.78\, H$, the adopted prescription coincides with $\overline{\mathrm{MS}}$ subtraction at this scale.

%%%%%%%%%%%%%%%%%%%%%%%%%%%%%%%%%%%%%%%%%%%%%%%%%%%%%%%%%%%%%%%%%%%%%%%%%

\section{Horizon-cutoff prescriptions and comparison}
\label{sec:approach_comparison}

The UV-divergent integrals in Eqs.~\eqref{rhoERaw} and \eqref{rhoBRaw} can be regulated by restricting the integrals to modes with momenta below a characteristic horizon scale. This prescription is motivated by the expectation that sub-horizon modes remain approximately vacuum-like, whereas super-horizon modes are significantly affected by the time-dependent kinetic coupling. Since the transition between these regimes is gradual, the choice of horizon scale is not unique. In this section, we examine five horizon-cutoff prescriptions and compare the resulting total gauge-field energy densities with the QFT-regularized result obtained in the previous section. The corresponding comparisons for the electric and magnetic energy densities separately are presented in Appendix~\ref{app:comparison-components}.

\subsection{Standard horizon cutoff}
\label{subsec:std}

A commonly used choice~\cite{Sobol:2020lec,Durrer:2023rhc} identifies the boundary by comparing the $k^2$ term with the coupling term in the magnetic mode equation \eqref{eqBRaw}:
\begin{equation}
\label{khA}
    k_h^{(\mathcal A)}
    =\sqrt{\left|\frac{f''}{f}\right|}.
\end{equation}
Here, $k_h^{(\mathcal A)}$ is the comoving momentum of the boundary mode at the time of evaluation. In this prescription, the same cutoff is applied to both energy densities, retaining only modes with $k\leq k_h^{(\mathcal A)}$:
\begin{equation}
\label{rhoEA}
    \rho_E^{(\mathcal A)}
    =\frac{1}{4\pi^2a^4}
    \int_0^{k_h^{(\mathcal A)}}\mathrm{d}k\,k^3
    |\mathcal D|^2,
\end{equation}
\begin{equation}
\label{rhoBA}
    \rho_B^{(\mathcal A)}
    =\frac{1}{4\pi^2a^4}
    \int_0^{k_h^{(\mathcal A)}}\mathrm{d}k\,k^3
    |\mathcal A|^2.
\end{equation}
The total energy density is defined as
\begin{equation}
\label{rhoGFA}
    \rho_{\mathrm{GF}}^{(\mathcal A)}
    =\rho_B^{(\mathcal A)}+\rho_E^{(\mathcal A)}.
\end{equation}

Under the slow-roll approximation in Eq.~\eqref{approx}, the cutoff becomes
\begin{equation}
    k_h^{(\mathcal A)}
    =aH\sqrt{|\gamma_1(\gamma_1+1)|}.
\end{equation}
Using the analytical mode functions from Sec.~\ref{subsec:analytic_mode_solutions}, we obtain
\begin{equation}
\label{rhoBA-integral}
    \rho_B^{(\mathcal A)}
    =\frac{H^4}{8\pi}
    \int_0^{\sqrt{|\gamma_1(\gamma_1+1)|}}
    \mathrm{d}x\,x^4
    \left|H_\alpha^{(1)}(x)\right|^2,
\end{equation}
and
\begin{equation}
\label{rhoEA-integral}
    \rho_E^{(\mathcal A)}
    =\frac{H^4}{8\pi}
    \int_0^{\sqrt{|\gamma_1(\gamma_1+1)|}}
    \mathrm{d}x\,x^4
    \left|H_\beta^{(1)}(x)\right|^2,
\end{equation}
where $x=k/(aH)$, $\alpha=|\gamma_1+\tfrac12|$, and
$\beta=|\gamma_1-\tfrac12|$.

\begin{figure}[t]
  \centering
  \includegraphics[width=0.95\columnwidth]{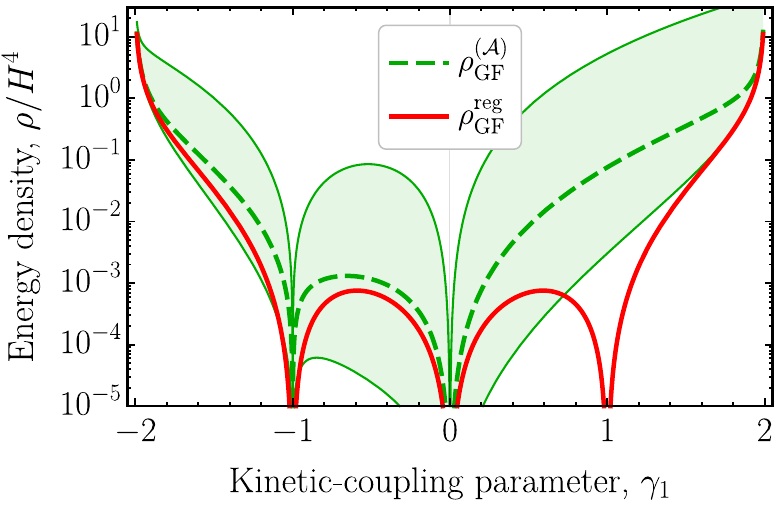}
  \caption{Comparison of the total gauge-field energy density obtained from QFT regularization (red solid curve) with that obtained using the standard horizon-cutoff prescription (green dashed curve), both normalized by $H^4$. The green shaded band illustrates the sensitivity of the horizon-cutoff result to the choice of cutoff scale, obtained by varying the horizon momentum by a factor of $\sqrt{10}$ upward and downward. This band provides an indicative estimate of the theoretical uncertainty associated with the horizon-cutoff prescription.}
  \label{fig:rhoGF_A}
\end{figure}

Since the cutoff is defined using the magnetic mode equation and then applied to both energy densities, this prescription does not preserve the electric--magnetic symmetry
$\rho_E(\gamma_1)=\rho_B(-\gamma_1)$ established in Sec.~\ref{subsec:analytic_mode_solutions}. In particular, at $\gamma_1=1$ it gives
\begin{equation}
    \rho_E^{(\mathcal A)}\big|_{\gamma_1=1}
    =\frac{H^4}{8\pi}
    \int_0^{\sqrt{2}}\mathrm{d}x\,x^4
    \left|H_{1/2}^{(1)}(x)\right|^2
    =\frac{H^4}{4\pi^2},
\end{equation}
rather than zero, as obtained for the regularized electric contribution. This illustrates a limitation of the standard horizon-cutoff prescription.

Figure~\ref{fig:rhoGF_A} compares the total gauge-field energy density obtained with the standard horizon-cutoff prescription (green dashed curve) with the regularized QFT result (red solid curve), both normalized by $H^4$. For $-2<\gamma_1<0$, the green dashed curve reproduces the zero at $\gamma_1=-1$, although its magnitude differs from that of the red curve over much of this interval. For $0<\gamma_1<2$, the horizon-cutoff result fails to reproduce the zero at $\gamma_1=1$ and substantially overestimates the total energy density. In Fig.~\ref{fig:rhoGF_A}, as in all figures in this section, the green shaded band illustrates the sensitivity of the horizon-cutoff result to the choice of cutoff scale. It is obtained by varying the nominal horizon momentum by a factor of $\sqrt{10}$ in either direction, so that the two extreme cutoff values differ by one order of magnitude. This band provides an indicative estimate of the theoretical uncertainty associated with the horizon-cutoff prescription.

\subsection{Separate horizon scales}
\label{subsec:separate}

The magnetic and electric mode equations, Eqs.~\eqref{eqBSim} and
\eqref{eqDSim}, respectively, suggest distinct horizon scales for the
two modes. While the magnetic horizon is defined by the coupling term
in the magnetic mode equation, as in Sec.~\ref{subsec:std}, Eq.~\eqref{khA},
the electric horizon follows from the effective coupling term in the
electric mode equation. We therefore define
\begin{equation}
\label{khD}
    k_h^{(\mathcal D)}
    =\sqrt{\left|2\left(\frac{f'}{f}\right)^2
    -\frac{f''}{f}\right|}.
\end{equation}
Under the slow-roll approximation in Eq.~\eqref{approx}, this becomes
\begin{equation}
\label{khD-slowroll}
    k_h^{(\mathcal D)}
    =aH\sqrt{|\gamma_1(\gamma_1-1)|}.
\end{equation}
In this prescription, each energy-density integral is
cut off at the horizon associated with its respective mode. 
Thus, the
magnetic energy density is $\rho_B^{(\mathcal A)}$, as defined in
Eq.~\eqref{rhoBA-integral}, while the electric energy density is
\begin{equation}
\label{rhoED}
    \rho_E^{(\mathcal D)}
    =\frac{H^4}{8\pi}
    \int_0^{\sqrt{|\gamma_1(\gamma_1-1)|}}
    \mathrm{d}x\,x^4
    \left|H_\beta^{(1)}(x)\right|^2.
\end{equation}
The total energy density for this separate-horizon prescription is
therefore
\begin{equation}
\label{rhoGFAplusD}
    \rho_{\mathrm{GF}}^{(\mathcal A+\mathcal D)}
    =\rho_B^{(\mathcal A)}+\rho_E^{(\mathcal D)}.
\end{equation}

\begin{figure}[t]
  \centering
  \includegraphics[width=0.95\columnwidth]{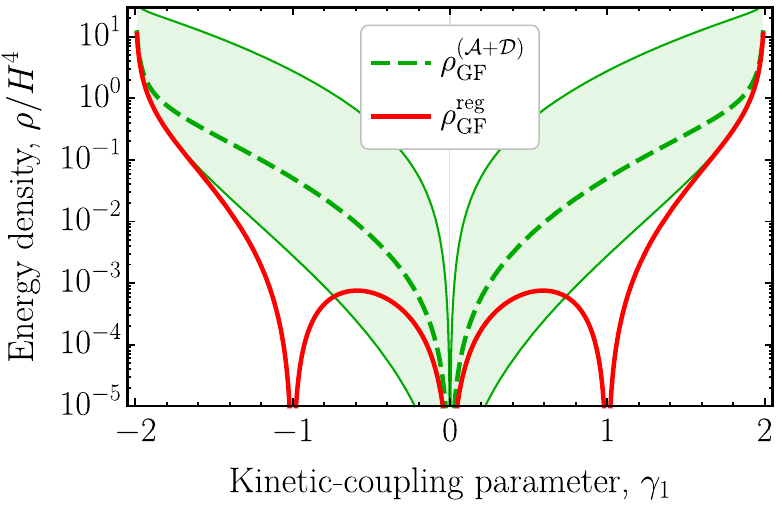}
  \caption{Comparison of the regularized QFT total gauge-field energy density, $\rho_{\mathrm{GF}}^{\mathrm{reg}}$, and the total energy density obtained using separate magnetic and electric horizon scales, $\rho_{\mathrm{GF}}^{(\mathcal A+\mathcal D)}$, normalized by $H^4$. The curve and shaded-band conventions are as described in the caption to Fig.~\ref{fig:rhoGF_A}.}
  \label{fig:rhoGF_AD}
\end{figure}

Because the magnetic and electric horizons transform into one another under $\gamma_1\to-\gamma_1$, this prescription preserves the electric--magnetic symmetry $\rho_E(\gamma_1)=\rho_B(-\gamma_1)$ established in Sec.~\ref{subsec:analytic_mode_solutions}. However, at $\gamma_1=-1$ the magnetic horizon vanishes, while at $\gamma_1=1$ the electric horizon vanishes; the other contribution remains nonzero at each point. Consequently, the total energy density fails to reproduce the zeros at $\gamma_1=\pm1$ found in the regularized QFT result. As shown in Fig.~\ref{fig:rhoGF_AD}, the prescription also yields a substantially larger total energy density over much of the plotted range. The shaded band has the same meaning as in Fig.~\ref{fig:rhoGF_A}.

\subsection{Hamiltonian horizon}
\label{subsec:Hamiltonian}

The magnetic and electric modes form a coupled Hamiltonian system, whose evolution is governed by the Hamiltonian equations \eqref{hamiltonEq}. The corresponding Hamiltonian matrix can be written as
\begin{equation}
\label{hamiltonian-matrix}
    \mathcal{H}
    =
    \begin{pmatrix}
        f'/f & k\\[4pt]
        -k & -f'/f
    \end{pmatrix}.
\end{equation}
Its eigenvalues are
\begin{equation}
\label{hamiltonian-eigenvalues}
    \lambda_{\pm}
    =\pm\sqrt{(f'/f)^2-k^2}.
\end{equation}
Depending on the sign of $\left(f'/f\right)^2-k^2$, the eigenvalues are either real or purely imaginary. The transition between these two regimes occurs at $k=|f'/f|$, which we use to define the Hamiltonian horizon:
\begin{equation}
\label{khH}
    k_h^{(\mathcal H)}
    =\left|\frac{f'}{f}\right|.
\end{equation}
Under the slow-roll approximation in Eq.~\eqref{approx}, this becomes
\begin{equation}
\label{khH-slowroll}
    k_h^{(\mathcal H)}
    =aH|\gamma_1|.
\end{equation}

We apply the Hamiltonian horizon cutoff to both the magnetic and
electric energy-density integrals. The corresponding energy densities
are
\begin{equation}
\label{rhoBH}
    \rho_B^{(\mathcal H)}
    =\frac{H^4}{8\pi}
    \int_0^{|\gamma_1|}\mathrm{d}x\,x^4
    \left|H_\alpha^{(1)}(x)\right|^2,
\end{equation}
and
\begin{equation}
\label{rhoEH}
    \rho_E^{(\mathcal H)}
    =\frac{H^4}{8\pi}
    \int_0^{|\gamma_1|}\mathrm{d}x\,x^4
    \left|H_\beta^{(1)}(x)\right|^2,
\end{equation}
where $\alpha=|\gamma_1+\tfrac12|$ and
$\beta=|\gamma_1-\tfrac12|$. The total energy density is then
\begin{equation}
\label{rhoGFH}
    \rho_{\mathrm{GF}}^{(\mathcal H)}
    =\rho_B^{(\mathcal H)}+\rho_E^{(\mathcal H)}.
\end{equation}

\begin{figure}[t]
  \centering
  \includegraphics[width=0.95\columnwidth]{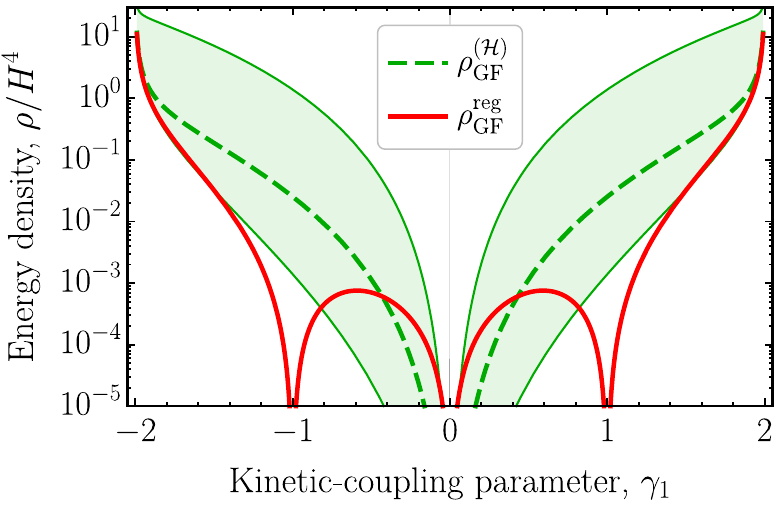}
  \caption{Comparison of the regularized QFT total gauge-field energy density, $\rho_{\mathrm{GF}}^{\mathrm{reg}}$, and the total energy density obtained using the Hamiltonian horizon prescription, $\rho_{\mathrm{GF}}^{(\mathcal H)}$, normalized by $H^4$. The curve and shaded-band conventions are as described in Fig.~\ref{fig:rhoGF_A}.}
  \label{fig:rhoGF_H}
\end{figure}

Since the same cutoff is applied to both sectors and is invariant under $\gamma_1\to-\gamma_1$, this prescription preserves the electric--magnetic symmetry $\rho_E(\gamma_1)=\rho_B(-\gamma_1)$. However, the Hamiltonian horizon remains nonzero at $\gamma_1=\pm1$, so the prescription fails to reproduce the zeros of the regularized total energy density at these points. Figure~\ref{fig:rhoGF_H} illustrates both features and shows that the horizon-cutoff result substantially overestimates the QFT-regularized energy density over much of the plotted range.

\subsection{Geometric-mean horizon}
\label{subsec:geometric-mean}

The magnetic and electric modes have distinct horizon scales,
$k_h^{(\mathcal A)}$ and $k_h^{(\mathcal D)}$, as discussed in
Sec.~\ref{subsec:separate}. To construct a single cutoff that
treats both sectors symmetrically, we define the geometric-mean horizon
as
\begin{equation}
\label{khgm}
    k_h^{(\mathrm{gm})}
    =\sqrt{k_h^{(\mathcal A)}k_h^{(\mathcal D)}}.
\end{equation}
Using the expressions for the two horizons under the slow-roll
approximation, this becomes
\begin{equation}
\label{khgm-slowroll}
    k_h^{(\mathrm{gm})}
    =aH\left(\gamma_1^2|\gamma_1^2-1|\right)^{1/4}.
\end{equation}

Applying this common cutoff to the magnetic and electric energy-density
integrals gives
\begin{equation}
\label{rhoBgm}
    \rho_B^{(\mathrm{gm})}
    =\frac{H^4}{8\pi}
    \int_0^{(\gamma_1^2|\gamma_1^2-1|)^{1/4}}
    \mathrm{d}x\,x^4
    \left|H_\alpha^{(1)}(x)\right|^2,
\end{equation}
and
\begin{equation}
\label{rhoEgm}
    \rho_E^{(\mathrm{gm})}
    =\frac{H^4}{8\pi}
    \int_0^{(\gamma_1^2|\gamma_1^2-1|)^{1/4}}
    \mathrm{d}x\,x^4
    \left|H_\beta^{(1)}(x)\right|^2,
\end{equation}
where $x=k/(aH)$,
$\alpha=|\gamma_1+\tfrac12|$, and
$\beta=|\gamma_1-\tfrac12|$. The total energy density is
\begin{equation}
\label{rhoGFgm}
    \rho_{\mathrm{GF}}^{(\mathrm{gm})}
    =\rho_B^{(\mathrm{gm})}+\rho_E^{(\mathrm{gm})}.
\end{equation}

The geometric-mean horizon is invariant under
$\gamma_1\to-\gamma_1$ and vanishes at $\gamma_1=\pm1$. Consequently,
this prescription preserves the electric--magnetic symmetry
$\rho_E(\gamma_1)=\rho_B(-\gamma_1)$ and reproduces the zeros of the
regularized total energy density at both $\gamma_1=-1$ and
$\gamma_1=1$.

\begin{figure}[t]
  \centering
  \includegraphics[width=0.95\columnwidth]{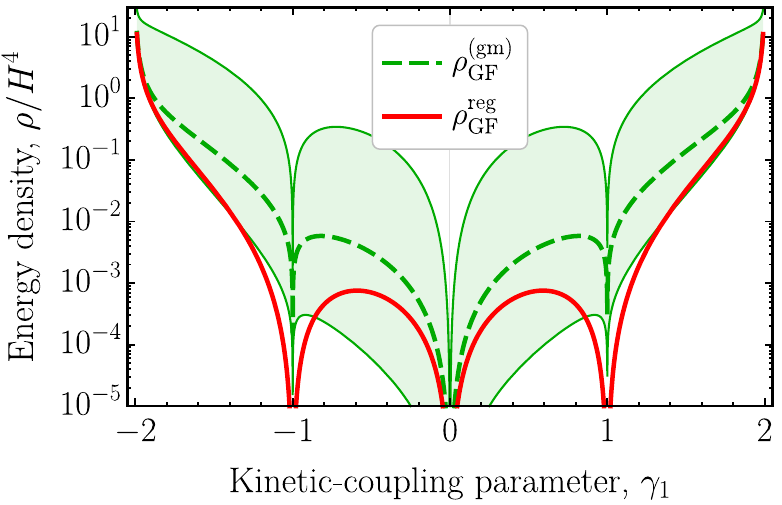}
  \caption{Comparison of the regularized QFT total gauge-field energy density, $\rho_{\mathrm{GF}}^{\mathrm{reg}}$, and the total energy density obtained using the geometric-mean horizon prescription, $\rho_{\mathrm{GF}}^{(\mathrm{gm})}$, normalized by $H^4$. The curve and shaded-band conventions are as described in Fig.~\ref{fig:rhoGF_A}.}
  \label{fig:rhoGF_gm}
\end{figure}

Figure~\ref{fig:rhoGF_gm} compares the total energy density obtained with the geometric-mean horizon prescription with the regularized QFT result. Although the geometric-mean horizon prescription reproduces the zeros at $\gamma_1=\pm 1$, its central prediction differs from the regularized QFT result over much of the plotted range. Nevertheless, the QFT result typically lies within the cutoff-variation band represented by the green shading in Fig.~\ref{fig:rhoGF_gm}.

\subsection{Minimal horizon}
\label{subsec:minimal}

The electric and magnetic modes are Hamiltonian dual variables and,
in the vacuum-dominated regime, should be treated on an equal footing.
However, as discussed in the preceding subsections, their associated
horizon scales differ, allowing the two modes to enter or leave this
regime at different times. To impose a common cutoff that does not
extend beyond either horizon, we define the minimal horizon as
\begin{equation}
\label{khmin}
    k_h^{(\min)}
    =\min\left(k_h^{(\mathcal A)},k_h^{(\mathcal D)}\right).
\end{equation}
Using the expressions for the magnetic and electric horizons derived in Sec.~\ref{subsec:separate}, we obtain
\begin{equation}
\label{khmin-slowroll}
    k_h^{(\min)}
    =aH\min\left(
    \sqrt{|\gamma_1(\gamma_1+1)|},
    \sqrt{|\gamma_1(\gamma_1-1)|}
    \right).
\end{equation}

A similar strategy, taking the smaller of two competing scales so that only modes below both are retained, was used in Refs.~\cite{Durrer:2024ibi,Teuscher:2025jhq}, where the minimum is taken over the two gauge-field modes in a convolution integral and over the horizon scales at two different times, respectively; here the same reasoning is applied to the electric and magnetic sectors.

Applying this cutoff to both sectors gives
\begin{equation}
\label{rhoBmin}
    \rho_B^{(\min)}
    =\frac{H^4}{8\pi}
    \int_0^{k_h^{(\min)}/(aH)}
    \mathrm{d}x\,x^4
    \left|H_\alpha^{(1)}(x)\right|^2,
\end{equation}
and
\begin{equation}
\label{rhoEmin}
    \rho_E^{(\min)}
    =\frac{H^4}{8\pi}
    \int_0^{k_h^{(\min)}/(aH)}
    \mathrm{d}x\,x^4
    \left|H_\beta^{(1)}(x)\right|^2,
\end{equation}
where $x=k/(aH)$,
$\alpha=|\gamma_1+\tfrac12|$, and
$\beta=|\gamma_1-\tfrac12|$. The total energy density is
\begin{equation}
\label{rhoGFmin}
    \rho_{\mathrm{GF}}^{(\min)}
    =\rho_B^{(\min)}+\rho_E^{(\min)}.
\end{equation}

\begin{figure}[b]
  \centering
  \includegraphics[width=0.95\columnwidth]{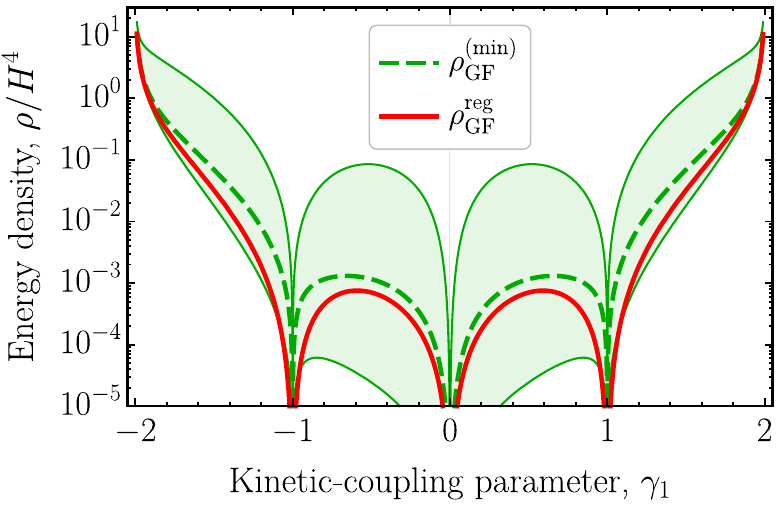}
  \caption{Comparison of the regularized QFT total gauge-field energy density, $\rho_{\mathrm{GF}}^{\mathrm{reg}}$, and the total energy density obtained using the minimal-horizon prescription, $\rho_{\mathrm{GF}}^{(\min)}$, normalized by $H^4$. The curve and shaded-band conventions are as described in Fig.~\ref{fig:rhoGF_A}.}
  \label{fig:rhoGF_min}
\end{figure}

Since the minimal horizon is invariant under
$\gamma_1\to-\gamma_1$, this prescription preserves the
electric--magnetic symmetry
$\rho_E(\gamma_1)=\rho_B(-\gamma_1)$. Moreover, the minimal horizon
vanishes at $\gamma_1=\pm1$, so the total energy density reproduces
the zeros of the regularized QFT result at these points.

Figure~\ref{fig:rhoGF_min} compares the total energy density obtained with the minimal-horizon prescription with the regularized QFT result.
Among the horizon prescriptions considered above, the minimal horizon provides the closest overall agreement with the QFT result in the plotted range. This agreement, however, does not extend to the individual magnetic and electric contributions (see Appendix~\ref{app:comparison-components}), for a structural reason: the horizon-cutoff integrals are nonnegative by construction, whereas the regularized components are not separately positive definite (Sec.~\ref{subsec:regularized-en-dens}). As $|\gamma_1|\to 2$, the infrared-enhanced component dominates, so that agreement in the total also implies agreement for this component, while the subdominant one remains prescription dependent.

%%%%%%%%%%%%%%%%%%%%%%%%%%%%%%%%%%%%%%%%%%%%%%%%%%%%%%%%%%%%%%%%%%%%%%%%%

\section{Conclusions}
\label{sec:conclusions}

In this paper, we have investigated the energy density associated with vacuum fluctuations of a $U(1)$ gauge field kinetically coupled to the inflaton during inflation. The corresponding energy-density integrals are UV divergent, requiring a prescription for extracting finite contributions. We have examined two approaches to this problem: (i)~conventional QFT regularization, using hard-cutoff and dimensional regularization, and (ii)~horizon-cutoff prescriptions, in which the integrals are restricted to modes below a characteristic horizon momentum. Within the zeroth-order slow-roll approximation, we derived analytical expressions for the gauge-field energy densities and compared the predictions of these approaches.
 
The QFT regularization analysis makes explicit the UV divergences that must be subtracted to obtain finite expressions. We have adopted a finite-part prescription that ensures nonnegativity of the coupling-induced total gauge-field energy density and respects the symmetry under the transformation $\gamma_1\rightarrow-\gamma_1$, accompanied by the interchange of the electric and magnetic sectors. This choice should be understood as a prescription for fixing the finite contribution within the approximation considered, rather than as a unique consequence of UV regularization, and the comparisons summarized below are made relative to it. A complete renormalization analysis would require a systematic treatment of the relevant local counterterms in the gravitational and matter sectors, which lies beyond the scope of the present study.
 
These results were compared with the conventional horizon-cutoff prescription, in which only modes with comoving momenta below a prescribed horizon scale are retained. Our analysis shows that the standard magnetic-horizon prescription can lead to substantial discrepancies relative to the QFT-based result. In particular, it does not preserve the symmetry under $\gamma_1\rightarrow-\gamma_1$, reproduces the vanishing of the regularized total energy density at $\gamma_1=-1$ but not at $\gamma_1=1$, and substantially overestimates the energy density for positive $\gamma_1$. These discrepancies indicate that, for weak gauge-field production, simply restricting the integrals to modes below a prescribed horizon scale need not provide a reliable estimate of the finite gauge-field energy density.
 
To investigate whether these shortcomings originate from the particular choice of the horizon, we considered several alternative prescriptions motivated by the mode equations and the Hamiltonian structure of the system. The separate-horizon prescription, which assigns distinct characteristic scales to the magnetic and electric modes, restores the electric--magnetic symmetry but does not reproduce the vanishing energy density at $\gamma_1=\pm 1$ and generally yields excessive energy densities. The Hamiltonian horizon also respects the symmetry, but retains significant discrepancies near $\gamma_1=\pm1$. The geometric-mean and minimal-horizon prescriptions improve the agreement with the QFT-regularized result at these points.
 
Among the prescriptions considered, and for the adopted finite-part prescription, the minimal horizon gives the closest overall agreement with the regularized QFT result for the total energy density over the parameter range examined, while for the geometric-mean prescription the regularized result typically lies within the estimated cutoff-variation band. A common feature of all five horizon-cutoff prescriptions is their improved agreement with the regularized QFT result in the regime of strong gauge-field production, as $|\gamma_1|$ approaches $2$. Although the prescriptions can differ substantially at intermediate values of $\gamma_1$, their relative deviation from the regularized result vanishes in this limit, while the uncertainty associated with variations of the horizon scale, represented by the shaded bands in the figures, becomes progressively smaller. This convergence reflects the dominance of infrared modes, whose contribution is insensitive to both the horizon scale and the finite-part prescription.
 
The individual electric and magnetic energy densities require separate consideration. As discussed in Sec.~\ref{sec:ultraviolet_regularization}, no choice of the finite-part constant makes both regularized components nonnegative throughout the parameter range, whereas the horizon-cutoff integrals are nonnegative by construction. Away from $|\gamma_1|\to 2$, agreement in the total energy density therefore does not imply that either component is reproduced reliably, as illustrated in Appendix~\ref{app:comparison-components}, and our assessment of the horizon-cutoff prescriptions focuses on the total energy density. This caveat is relevant for applications in which the two sectors play different roles, such as inflationary magnetogenesis, where the magnetic sector characterizes the generated fields while the backreaction on the inflaton is governed by the difference $\rho_E-\rho_B$ [see Eq.~\eqref{inflatonEq}]; predictions for the relative importance of the two sectors can depend sensitively on the prescription used to extract the finite energy density.
 
Several limitations of the present analysis should be acknowledged. First, our calculations employ a simplified inflationary background and an approximately constant kinetic-coupling parameter. This approximation does not capture the full dependence on background derivatives and curvature that would generally enter a systematic determination of the relevant counterterms, and the residual finite-part ambiguity need not reduce to the single constant $C$ considered here. Consequently, the finite expressions obtained here should be regarded as a controlled comparison of regularization prescriptions within the adopted approximation, rather than as a complete renormalization of the energy--momentum tensor in a general time-dependent background.
 
Second, the regularized expressions cannot be directly substituted into the background equations to determine the backreaction of the produced gauge fields without a more systematic treatment of renormalization and the associated counterterms.
 
A natural direction for our future work is therefore to extend the analysis beyond the leading slow-roll approximation and investigate the renormalization of the full energy--momentum tensor in a self-consistent inflationary background. Such an extension would make it possible to establish the role of the finite renormalization conditions more systematically and to determine how the regularized energy density evolves when the kinetic coupling varies appreciably with time. It would also provide a firmer basis for incorporating gauge-field backreaction into the inflaton and Friedmann equations. Finally, applying the resulting framework to concrete models of inflationary magnetogenesis and related scenarios would allow the impact of the regularization procedure on their phenomenological predictions to be assessed quantitatively.

%%%%%%%%%%%%%%%%%%%%%%%%%%%%%%%%%%%%%%%%%%%%%%%%%%%%%%%%%%%%%%%%%%%%%%%%%

%%%%%%%%%%%%% Acknowledgements %%%%%%%%%%%%%%%%%%%%%%%%%%%%%%%%%%%%%%%%%%
\acknowledgments

S.~I.~V. is grateful to Prof. Marc Vanderhaeghen and Dr. Vladimir Pascalutsa  for their support, useful discussions, and kind hospitality at the Johannes Gutenberg University Mainz, Germany, where most of this work was done.
The work of O.~O.~S. was funded by the Deutsche Forschungsgemeinschaft (DFG, German Research Foundation) --- project number 565969529.
At the initial stage of this project, the work of O.~O.~S. has received funding through the SAFE --- Supporting At-Risk Researchers with Fellowships in Europe project, which is funded by the European Union. Views and opinions expressed are, however, those of the authors only and do not necessarily reflect those of the European Union or the European Research Executive Agency (REA). Neither the European Union nor the granting authority can be held responsible for them.
The work was partially supported by the project ``Search for dark matter and particles beyond the Standard Model'' of the Ministry of Education and Science of
Ukraine 25BF051-01.

%%%%%%%%%%%%%%%%%%%%%%%%%%%%%%%%%%%%%%%%%%%%%%%%%%%%%%%%%%%%%%%%%%%%%%%%%

%\paragraph{Note added.} This is also a good position for notes added after the paper has been written.

% %%%%%%%%%%%%%%%% APPENDICES %%%%%%%%%%%%%%%%%%%%%%%%%%%%%%%%%%%%%%%%%%%%%

\appendix
\begin{widetext}
\section{Hankel-function integrals and asymptotic expansion}
\label{app:Hankel-integrals}

In order to evaluate the magnetic energy density in hard-cutoff regularization, Eq.~\eqref{rhoBLambda}, we need to compute the following integral:
\begin{equation}
\label{integralRaw}
    I(x;\alpha) = \int_{0}^{x}\text{d}t\,t^4\left|H^{(1)}_{\alpha}(t)\right|^2\,,
\end{equation}
where $\alpha$ is a real parameter, which, in the model under consideration, can vary in the range $|\alpha|<\frac{5}{2}$. Since $\alpha$ is real, we expand the squared modulus of the Hankel function as:
\begin{equation}
\label{appA:SquaredModuleHankelsExpansion}
    \left|H_\alpha^{(1)}(t)\right|^2 =
    \frac{J_\alpha^2(t)+J_{-\alpha}^2(t)
    -2\cos(\pi\alpha)J_\alpha(t)J_{-\alpha}(t)}{\sin^2(\pi\alpha)},
\end{equation}
Thus, the integral in Eq.~\eqref{integralRaw} can be decomposed into three integrals involving products of Bessel functions. For these integrals we apply Eq.~(1.8.3.1) of Ref.~\cite{prudnikov}, which has the form
%\begin{widetext}
\begin{multline}\label{appA:UpperBoundBesselIntegral}
    \mathcal{I}(\lambda,\,\mu,\,\nu;\,x)=\int_0^x\,t^\lambda J_\mu(t)J_\nu(t)\,\text{d}t=\frac{x^{\lambda+\mu+\nu+1}}{2^{\lambda+\mu}(\lambda+\mu+\nu+1)\Gamma(\mu+1)\Gamma(\nu+1)}\\ \times{}_3 F_4\left(\frac{\mu+\nu+1}{2},\,\frac{\mu+\nu+2}{2}\,,\frac{\lambda+\mu+\nu+1}{2};\,\mu+1,\,\nu+1,\,\mu+\nu+1,\,\frac{\lambda+\mu+\nu+3}{2};\,-x^2\right)\,,\\\quad\text{Re}(\lambda+\mu+\nu)>-1\,.
\end{multline}
Here, ${}_{3}F_{4}$ denotes a generalized hypergeometric function. Applying the expression \eqref{appA:UpperBoundBesselIntegral}  with $\lambda=4$ and corresponding $\mu$ and $\nu$, we obtain
\begin{multline}
I(x;\alpha) =\frac{\csc^2(\pi\alpha){}_2F_{3}\left(\alpha+\frac{1}{2},\alpha+\frac{5}{2};\alpha+1,\alpha+\frac{7}{2},2\alpha+1;-x^2\right)}{4^{\alpha}(5+2\alpha)\Gamma(\alpha+1)^2}x^{5+2\alpha}\\
+\frac{\csc^2(\pi\alpha){}_2F_{3}\left(\frac{1}{2}-\alpha,\frac{5}{2}-\alpha;1-2\alpha,1-\alpha,\frac{7}{2}-\alpha;-x^2\right)}{4^{-\alpha}(5-2\alpha)\Gamma(1 -\alpha)^2}x^{5-2\alpha}\\
-\frac{2\cot(\pi\alpha){}_{2}F_{3}\left(\frac{1}{2},\frac{5}{2};\frac{7}{2},1-\alpha,1+\alpha;-x^2\right)}{5\pi\alpha}x^5\,.
\label{integral-Hankel-finite-upper-limit}
\end{multline}
%\end{widetext}
where we took into account that after canceling the common upper and lower parameter, the generalized hypergeometric function ${}_3 F_4$ reduces to ${}_2 F_3$.
\begin{comment}
\begin{equation}
    {}_{p+1}F_{q+1}\left(a_1\,,\dots,\,a_p,\,c;\,b_1\,,\dots,\,b_q\,,c;\,z\right)={}_{p}F_{q}\left(a_1\,,\dots,\,a_p,\,\,b_1\,,\dots,\,b_q\,;\,z\right)\,.
\end{equation}
\end{comment}
The asymptotic expansion of the expression \eqref{integral-Hankel-finite-upper-limit} for $x\to\infty$ is
\begin{multline}
\label{int-Hankel-finite-upper-limit-asympt}
    I(x;\alpha)=\frac{x^4}{2\pi}+\frac{(4\alpha^2-1)}{8\pi}x^2+\frac{3(4\alpha^2-1)(4\alpha^2-9)}{64\pi}\ln{(x)}\\
    +\frac{(4\alpha^2-1)(4\alpha^2-9)}{256\pi}\Bigg[-6\psi^{(0)}\left(-\alpha-\frac{3}{2}\right)-6\psi^{(0)}\left(\alpha-\frac{3}{2}\right)-7+6\ln{4}\Bigg]+\mathcal{O}\left(\frac{1}{x}\right)\,.
\end{multline}
This expression is then used in the main text with the substitution $\alpha = \left|\gamma_1 + \frac{1}{2}\right|$ to obtain Eq.~\eqref{rhoBGammaRenorm}.

In order to compute the magnetic energy density in dimensional regularization, Eq.~\eqref{rhoBdimreg-integral}, we need to evaluate an integral of the form
\begin{equation}
    I(d,x;\alpha)=\frac{2^{-d} x^{4-d} \pi^{\frac{3}{2}-\frac{d}{2}}}{\Gamma \left(\frac{d-1}{2}\right)}\int\limits_0^{+\infty} \mathrm{d}t\, t^d\,\left|H^{(1)}_{\alpha}(t)\right|^2\,.
\end{equation}
First, using Eq.~\eqref{appA:SquaredModuleHankelsExpansion} and then applying the integral formula in Eq.~(6.574.2) of Ref.~\cite{GradshteinRyzhik:2007handbook}, which is of the form
\begin{align}
\mathcal{I}(\lambda;\mu,\nu)=\int_0^\infty J_\nu(t)J_\mu(t)t^{-\lambda}\,dt
=
\frac{
\Gamma(\lambda)
\Gamma\left(\frac{\nu+\mu-\lambda+1}{2}\right)
}{
2^\lambda
\Gamma\left(\frac{-\nu+\mu+\lambda+1}{2}\right)
\Gamma\left(\frac{\nu+\mu+\lambda+1}{2}\right)
\Gamma\left(\frac{\nu-\mu+\lambda+1}{2}\right)
}\,,
\end{align}
we obtain
%\begin{widetext}
\begin{multline}
\label{app-integral-dimreg-full}
    I(d,x;\alpha) = \frac{2^{-d} x^{4-d} \pi^{\frac{3}{2}-\frac{d}{2}}}{\Gamma \left(\frac{d-1}{2}\right)}\frac{\mathcal{I}(-d;\alpha,\alpha)+\mathcal{I}(-d;-\alpha,-\alpha)-2\cos(\pi\alpha)\mathcal{I}(-d;\alpha,-\alpha)}{\sin^2(\pi\alpha)}\\
    = \frac{x^{4-d}\pi^{\frac{3}{2}-\frac{d}{2}} (d-1)\Gamma (-d) \cos (\pi  \alpha )}{\Gamma \left(\frac{1-d}{2}\right) \Gamma \left(\frac{1-d}{2}+\alpha \right) \Gamma \left(\frac{1-d}{2}-\alpha \right) \cos \left[\pi  \left(\frac{d}{2} + \alpha\right)\right] \cos \left[\pi  \left(\frac{d}{2} - \alpha \right)\right]}\, .
\end{multline}
Now, setting $d=4-\epsilon$ and expanding the expression above up to and including $\mathcal{O}(\epsilon^0)$, we obtain
\begin{equation}
\label{app-integral-dimreg-expansion}
    I(4-\epsilon,x;\alpha) = \frac{3(4\alpha^2-1)(4\alpha^2-9)}{1024 \pi ^2} \left[\frac{2}{\epsilon}+\ln(4\pi)-\gamma_E+\frac{5}{6}+\ln(x^2)-\psi ^{(0)}\left(-\alpha -\frac{3}{2}\right)-\psi ^{(0)}\left(\alpha -\frac{3}{2}\right)\right] + \mathcal{O}(\epsilon).
\end{equation}
This expression is then used in the main text with the substitution $\alpha = \left|\gamma_1 + \frac{1}{2}\right|$ to obtain Eq.~\eqref{rhoBdimreg-expansion}.
%\end{widetext}

\section{Electric and magnetic energy densities under horizon-cutoff prescriptions}
\label{app:comparison-components}

In this appendix, we compare the electric and magnetic energy densities obtained using the five horizon-cutoff prescriptions introduced in Sec.~\ref{sec:approach_comparison} with their regularized QFT counterparts. While the total energy density provides an overall measure of the agreement between the two approaches, the separate components reveal differences that can be obscured when they are summed. In particular, a prescription that reproduces the total energy density reasonably well need not accurately capture the electric and magnetic contributions individually. The panels in Fig.~\ref{fig:rhoEB_comparison} show these component-wise comparisons for all five prescriptions, allowing their performance and their preservation of the electric--magnetic symmetry to be assessed separately.
\end{widetext}

\begin{figure*}[t]
  \centering
  \includegraphics[width=0.8\textwidth]{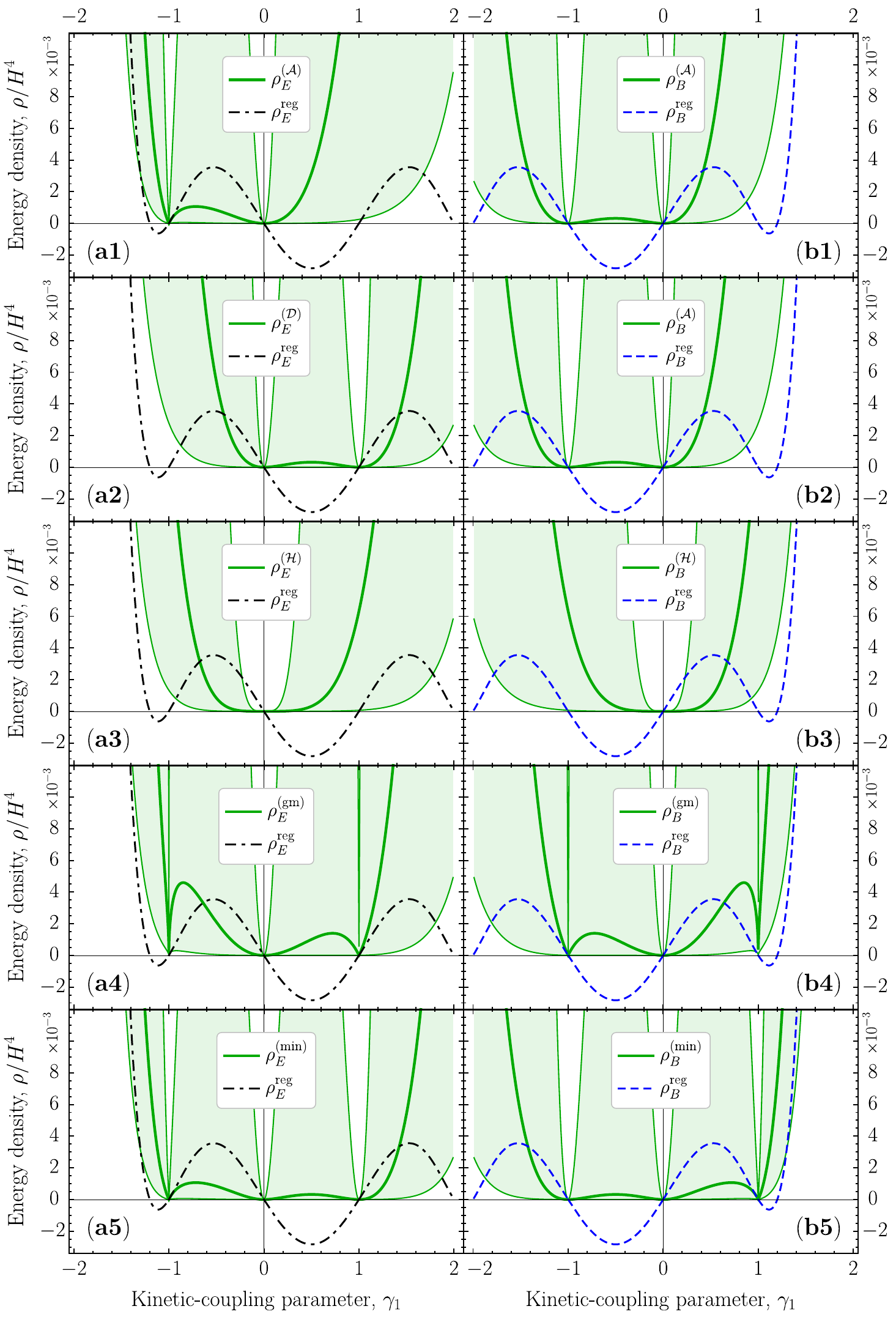}
  \caption{Comparison of the QFT regularized electric and magnetic energy densities with those obtained using the five horizon-cutoff prescriptions discussed in Sec.~\ref{sec:approach_comparison}. The left column (panels a1--a5) shows the electric energy density, while the right column (panels b1--b5) shows the magnetic energy density (panels b1 and b2 coincide by construction). From top to bottom, the rows correspond to the standard horizon, separate horizon scales, Hamiltonian horizon, geometric-mean horizon, and minimal horizon prescriptions, respectively. All energy densities are normalized by $H^4$. The black dashed--dotted curves and blue dashed curves represent the regularized QFT result for the electric and magnetic energy densities, respectively. The green solid curves represent the corresponding horizon-cutoff result, and the green shaded band indicates the sensitivity to variations of the cutoff scale by a factor of $\sqrt{10}$ in either direction.}
  \label{fig:rhoEB_comparison}
\end{figure*}

%%%%%%%%%%%%%%% Bibliography %%%%%%%%%%%%%%%%%%%%%%%%%%%%%%%%%%%%%%%%%%%%
% \bibliographystyle{JHEP}
% \bibliography{references.bib}

%\providecommand{\href}[2]{#2}\begingroup

%\endgroup

%%%%%%%%%%%%%%%%%%%%%%%%%%%%%%%%%%%%%%%%%%%%%%%%%%%%%%%%%%%%%%%%%%%%%%%%%

\end{document}